\documentclass[twoside,11pt] {article}  

\usepackage{amsmath,amssymb}

\newtheorem{twr}{Theorem}

\newcommand{\ud}{\mathrm{d}}

\begin{document}           

\title{NONCOMMUTATIVE SPACETIME AND 
QUANTUM MECHANICS}  
\author{Jaros{\l}aw Wawrzyckii\footnote{Electronic address: 
Jaroslaw.Wawrzycki@ifj.edu.pl}
\\Institute of Nuclear Physics, ul. Radzikowskiego 152, 
\\31-342 Krak\'ow, Poland}         

\maketitle                 

\pagestyle{myheadings}
\thispagestyle{plain}        
\markboth{\scriptsize JAROS{\l}AW WAWRZYCKI}{\scriptsize NONCOMMUTATIVE SPACETIME AND 
QUANTUM MECHANICS } 
\setcounter{page}{1}

%
%
%

\begin{abstract}
In this paper we will analyze the status of gauge freedom in quantum 
mechanics (QM) and quantum field theory (QFT). Along with this analysis
comparison with ordinary QFT will be given. We will show how the gauge 
freedom problem is connected with the spacetime coordinates status ---
the very point at which the difficulties of QM begin. A natural solution
of the above mentioned problem will be proposed in which we give a slightly 
more general form of QM and QFT (in comparison to the ordinary QFT) 
with noncommutative structure of spacetime playing fundamental role in it. We 
achieve it by reinterpretation of the Bohr's complementarity principle
on the one hand and by incorporation of our gauge freedom analysis
on the other. We will present a generalization of the Bargmann's
theory of exponents of ray representations.
It will be given an example involving time dependent gauge 
freedom describing non-relativistic quantum particle in nonrelarivistic 
gravitational field. In this example we infer the most general Schr\"odinger
equation and prove equality of the (passive) inertial and the gravitational  
masses of quantum particle. 
\end{abstract}

\section{General Introduction}

Probably it will be helpful to give a brief outline of this paper 
providing the general line of reasoning. For details, however, the reader 
must consult the foregoing sections. 

Of late there has been proposed a reformulation of the standard QM      
({\itshape J. Wawrzycki, math-ph/0301005; Comm. Math. Phys., to appear}),
which is slightly more general in comparison to the ordinary form
of the theory. This reformulation has emerged in a natural way in description
of a quantum particle in the non-relativistic gravitational field with time
dependent gauge freedom. Remember, please, that the states of a physical 
system do not correspond bi-uniquely to unit vectors $\phi$ of the respective
Hilbert space $\mathcal{H}$ but to the rays, say $\boldsymbol{\phi} 
= \{e^{i\xi} \phi\}$ in the QM and QFT, where $\xi$ is an arbitrary 
real number.\footnote{Let us recall that all relevant
information carried by $\phi$ is contained in the set of numbers 
\begin{displaymath}
\frac{\vert(\phi, \varphi)\vert^{2}}{(\phi,\phi)(\varphi, \varphi)},   
\end{displaymath}
where $(\phi, \varphi)$ is the scalar product of the Hilbert space 
$\mathcal{H}$. As such $\phi$ and $e^{i\xi} \phi$ are equivalent containing
exactly the same information.} Observe now, please, that in ordinary 
(non-relativistic) QM, when using Schr\"odinger picture, one can go 
considerably further with this observation. Namely, two Schr\"odinger
waves $\psi$ and $e^{i\xi(t)}\psi$ are indistinguishable even when
$\xi$ depends on time, but one have to assume simultaneously that  
Schr\"odinger wave equation possess a time dependent gauge freedom.
Let us recall that the integral defining the scalar product is over the
space coordinates and one can take a time dependent factor over the 
integral sign. After this, however, the Schr\"odinger wave functions should 
constitute the appropriate cross sections of a Hilbert bundle 
$\mathcal{\mathbb{R}}\triangle \mathcal{H}$ over time $\mathbb{R}$. 
The representations $T_{r}$ of a covariance as well as a symmetry
groups act in $\mathcal{\mathbb{R}}\triangle \mathcal{H}$ and their exponents
$\xi$ in the formula 
\begin{equation}\label{rayrep}
T_{r}T_{s} = e^{i\xi(r,s,t)}T_{rs},
\end{equation}
do depend on time $t \in \mathbb{R}$. Thus, at first sight the natural 
assumption that two Schr\"odinger waves differing by a time dependent phase
are equivalent leads to a rather strange construction, namely, the Hilbert bundle --- 
an object much more involved then the Hilbert space itself. One can prove, however, that
in the non-relativistic Galilean invariant theory, this more general assumption
leads exactly to the ordinary QM. The whole structure degenerates due to the 
specific structure of the Galilean group. Moreover, in the less trivial case of 
a quantum particle in non-relativistic gravitational field, when the time dependent 
gauge freedom is indispensable, the results are quite interesting. Namely, one can
infer the most general wave equation for that particle and prove equality of the 
inertial and gravitational mass, the results confirmed by experiments! In the last 
case the Milne group plays the role of the Galilean group. 
 
This non-relativistic generalization possesses also a very natural relativistic 
extension which can be incorporated within QFT rather then QM. In QFT one can 
still go a step further along with this line of generalizing the quantum 
mechanical principles. Remember, please, that in the Fock construction the 
Fourier components of the classical field constitute the arguments of the wave function.
Anyway the arguments have nothing to do with ordinary spacetime coordinates. In 
other words the spacetime coordinates are mere parameters or the so-called c-numbers
in Heisenberg canonical field quantization --- just like the time in ordinary 
non-relativistic QM. One should, thus, assume the two wave functions to be equivalent 
whenever they differ by a spacetime dependent wave function. But, when treating
this assumption seriously, the wave functions should constitute the appropriate 
cross sections of a Hilbert bundle $\mathcal{M}\triangle \mathcal{H}$ over 
spacetime $\mathcal{M}$. Accordingly the representations $T_{r}$ of covariance 
or symmetry groups possess spacetime dependent exponents $\xi = \xi(r,s,p)$:
\begin{displaymath}
T_{r}T_{s} = e^{i\xi(r,s,p)}T_{rs},
\end{displaymath}
with $p\in \mathcal{M}$ ({\itshape J. Wawrzycki, math-ph/0301005; Comm. Math. 
Phys., to appear}). 

In particular one is forced to extend the ordinary classification theory of exponents
$\xi$ of representations acting in ordinary Hilbert space so as to embrace the above 
case of representation acting in a Hilbert bundle with spacetime dependent $\xi$.   
It can be viewed as a generalization of the Bargmann's theory ({\itshape V. Bargmann, 
Ann. Math {\bf 54}, 1, 1954}) of exponents of ray representations acting in ordinary 
Hilbert space  with spacetime-independent $\xi$. 
 
The fact that the simpler form of the theory with time dependent gauge freedom 
gives the correct form of the Schr\"odinger equation thought the author to treat 
seriously also the generalization with spacetime dependent gauge freedom. The natural
question emerges if one can find a simple connection of our generalization (those with
spacetime dependent gauge freedom) to the ordinary QFT, for example to QED and
if the connection is so transparent as in the case of time dependent gauge freedom.
At this place one have to note that in particular any representation $T_{r}$ with
spacetime dependent exponent $\xi = \xi(r,s,p)$ in (\ref{rayrep}) makes any sense if the 
representation of the algebra of Canonical Commutation Relations (CCR) is {\bf reducible} 
and does possesses a {\bf nontrivial} center. The diagonal algebra over which the above 
representation of CCR algebra can be decomposed into irreducible components corresponds
to the classical spacetime. How one can interpret this strange result along with the 
ordinary QFT in which, according to ``Wightman's axioms'' ({\itshape A. S. Wightman
and L. G{\aa}rding, ``Fields as Operator Valued Distributions in Relativistic Quantum'' 
Theory, Ark. Fys. {\bf 23}, No. 13, 1964}), the algebra generated by quantum field 
operators (``smeared with appropriate test functions'') is {\bf irreducible}?

Answer to this question is by no means trivial. In particular the situation is much 
less trivial then in the case of the theory with time dependent gauge freedom in 
non-relativistic QM. In our opinion one is forced to reconsider the fundamental principles 
of QM and QFT in answering the question. Of late the long-lasting dispute concerning
the interpretation of QM is at its renaissance again. We will not go into detail of this 
dispute but we feel that something is missing in the standard picture and at least some 
points of the criticism are justified. We assume that the standard QM describes correctly
the situation in which we have an atomic system measured with a macroscopic apparatus.
In the standard QM the Bohr-Heisenberg cut between the system and the apparatus         
may range between them, and it is irrelevant how big is the system. It is important only
that the cut is ``placed'' somewhere between the system and the apparatus (in this sense 
the Bohr-Heisenberg cut does not form any ``normal'' real physical boundary). This 
unavoidably implies that a macroscopic body can be in a (quantum) superposition of states
with macroscopically different parameters like the center-of-mass position. On the other 
hand, so long as ``big'' and ``small'' are merely relative concepts, it is no help to  
one who wishes to account for the ultimate structure of matter. We agree with Dirac, that
QM should be endowed with ideas in such a way as to give an absolute meaning to the
words ``small'' and ``big''. In the common opinion QM is the theory which incorporates
the idea of absolute scale of action --- the Planck's quantum of action $\hslash$.
Paradoxically, in ordinary QM the only idea, namely that of Bohr-Heisenberg cut, 
which separates the ``small'' from the ``big'' is purely relative such that we lose
the possibility of introducing the absolute scale mentioned above. This is the element
which in our opinion is missing in the standard interpretation of QM. Thus, we assume that 
QM with the standard interpretation works good but only within some limits, where the 
observed system involves only a few quantum particles whereas the apparatus constitute
a macroscopic body consisting of an enormous number of quantum particles. But then the 
QFT is the appropriate scheme within which the systems consisting of many particles
are naturally incorporated. Therefore, one has to be careful in applying the QM principles 
in QFT when the number of particles is too big. The application is justified only if 
one works within the neighborhood of the vacuum state --- which does actually take 
place in practice e.g. when evaluating the {\itshape Lamb shift} or the {\itshape 
anomalous magnetic moment}. For the states far removed from the vacuum one has to
preserve an open mind on the ordinary QM principles. We maintain that our generalization
of QFT is applicable if the number of particles in the system is large, just in
opposite to the ordinary QFT applicable when the number is small. Indeed. The 
``Fock space'' $\mathcal{H}_{F}$ is the direct sum of all $N$-particle spaces
$\mathcal{H}_{N} =\mathcal{H}_{1}^{\otimes_{S} N}$ (resp. $\mathcal{H}_{N} =
\mathcal{H}_{1}^{\otimes_{A} N}$) i.e. the symmetrized $N$-fold products of 
the one-particle Hilbert space $\mathcal{H}_{1}$ (resp. anti-symmetrized 
$N$-fold products):
\begin{displaymath}
\mathcal{H}_{F} = {\bigoplus}_{N=0}^{\infty} \mathcal{H}_{N}.
\end{displaymath}       
A state vector $\psi$ in $\mathcal{H}_{F}$ is an infinite hierarchy of symmetric 
(resp. anti-symmetric) wave functions
\begin{displaymath}
\psi = \left( \begin{array}{ll}
c & \\
\psi_{1}(x) & \\
\psi_{2}(x_{1},x_{2}) & \\
\ldots &
\end{array} \right)
\end{displaymath}
where we have written the argument $x$ to denote both position and spin component.
$\psi_{N}$ is the probability amplitude for finding just $N$ particles and those 
in specified configuration. The complex number $c$ is the probability amplitude
to find the vacuum. It is important to note that the argument $x$ in one particle 
state $\psi_{1}$ contains the space coordinates whereas $x_{1}, \ldots , x_{N}$ in
the $N$-particle state have nothing to do with spacetime coordinates being rather
the configuration coordinates. In the case of the free field when the number of
particles $N$ is conserved our argument that $\psi$ and $e^{i\xi(p)}\psi$ are equivalent
is valid (provided $N>1$). Remember, please, that $p$ stands for spacetime point. 
In this case the probability amplitude to find a number of particles different 
from $N$ is zero (in particular $\psi_{1} = 0$).\footnote{Probably it will be 
helpful to recall the formula for the scalar product in
$\mathcal{H}_{F}$ 
\begin{displaymath}
(\psi,\psi') = c^{*}c' + \int_{\mathbb{R}^{3}}\psi_{1}^{*}\psi_{1}' \, d^{3}x  + 
\int_{\mathbb{R}^{2\times 3}}\psi_{2}^{*}\psi_{2}' \, d^{3}x_{1}d^{3}x_{2} + \ldots
\end{displaymath}  
to see the equivalence of $\psi$ and $e^{i\xi(p)}\psi$, compare the preceding 
footnote.} However, relativistic QFT with interaction is a theory describing the phenomenon
of creation and annihilation of particles as indispensable effects so as the total number
of particles cannot be conserved. Thus, one cannot exclude the amplitude $\psi_{1}$ 
when working around the vacuum state. This one-particle amplitude is negligible 
only if the total number of particles is large. In this case, therefore 
$\vert(e^{i\xi(p)}\psi,\psi')\vert = \vert e^{i\xi(p)}(\psi,\psi')\vert = 
\vert(\psi,\psi')\vert $ so that $\psi$ and $e^{i\xi(p)}\psi$ are equivalent;
please, compare the first two footnotes --- one can take the factor $e^{\xi(p)}$
over the argument of the inner product in this case. This shows that our 
generalization becomes adequate when the system consists of an appropriately large 
number of particles, just in the situation when the ordinary QFT is expected to 
be somewhat misleading. 

Yet the (more) ultimate general theory, the scientist should intend to find, is 
expected to be adequate to account for systems of a few particles as well as 
systems consisting of an enormous number of them. We are certain that the ordinary
QFT laws\footnote{The term ``ordinary'' means here the ordinary QM applied to 
system with infinite number of degrees of freedom.} works perfectly in these cases
of small values $N$ of particles in the system. Should it be possible at all to 
obtain a theory which can give correct answers also in the cases of small 
$N$-values? In the cases of small $N$ the ordinary QFT could act as the code-book
--- namely, one should pick up those general laws or theorems of QFT which can 
be formulated in terms independent of the actual number of particles.  
As such theorem which should guide us in 
further research we take the miracle that: {\bf 1) Canonical quantization of a free 
scalar field leads to Fock space and an interpretation of states in terms of 
particle configurations}, so it explains the wave-particle duality lying at the 
roots of QM. This theorem contains the general kinematical information 
that the states of quantized field serve as configuration space for quantum
particles which may occupy them. As such this fact does not depend directly 
on the actual value of $N$. On the other hand, when the number $N$ of particles 
is large the classical theory is correct and may serve as a guide in this
regime exactly as QFT does when $N$ is small, provided the laws we pick up
to guide us do not depend on the number of degrees of freedom\footnote{Remember, 
please, that the number of particles $N$ in quantum theory corresponds to to the 
number of degrees of freedom in classical theory, where the degree of freedom 
is in the sense of the Lagrange-Hamilton theory. The laws 
serving us as guiding receipt should not depend on $N$ as we have mentioned 
earlier. Therefore, those laws --- if taken from classical theory --- should not depend 
on the number of Lagrange degrees of freedom.}, especially they should not depend 
on the fact if there is infinite number of them as in classical field 
or finite as e.g. for a rigid classical body. As the second guiding theorem we take the
following theorem: {\bf 2) The configuration space for the classical body consists of various 
space\footnote{Here ``space'' stands for an 
appropriate space-like Cauchy section of (classical) spacetime.} positions the body 
or its parts may occupy} --- a fact which seems to be independent of the actual 
number of degrees of freedom involved, i.e. independent of the number of independent
parts. Next, we should observe that the algebraic formulation of QM and QFT is an 
approach in which the Hilbert space plays a secondary role and by this fact the 
number of particles $N$ as well as the number of degrees of freedom plays a very 
indirect role in it. Thus, the approach is the one we should work with in our research.    
Comparing now our guiding theorems 1) and 2) one can infer that the
(noncommutative) quantum algebra of quantum field operators\footnote{More precisely
we have to construct first the specific representation of the algebra with the vacuum
plying the role of the cyclic vector of the representation. Then, the 
the states of the Hilbert space of the representation constitute the analogue
of classical space positions.} plays the role 
of noncommutative space for quantum particles, just like the classical space 
(i.e. space-like surface of spacetime) does for a classical body. In this way 
we arrive at the result that the space the quantum particles live in is a
noncommutative space (in the sense of {\itshape A. Connes, Noncommutative Geometry, 
Acad. Press 1994}) with the noncommutative algebra of quantum field operators 
corresponding to the noncommutative space. 

Still, it would be much better if we were able to find the counterpart of the 
whole classical spacetime and not only the counterpart of space.\footnote{The
notion of a noncommutative space at a particular classical time is rather
a strange mixture of classical and noncommutative properties which cannot
serve as a generally valid universal concept.} For this purpose
let us note that the points of classical spacetime go into play when considering
the time evolution of space position of the classical body. {\itshape Per 
analogiam}, if we are to have any hope to find the quantum counterpart of spacetime 
we must use the quantum substitute for the classical space position evolution.
The state in the Fock space $\mathcal{H}_{F}$ in the Heisenberg picture (which we 
have used above) does not contain any information about time evolution. 
One could suppose that the simple replacement of the Heisenberg picture with
the Schr\"odinger picture immediately resolves the problem. Unfortunately this
is by no means the case. One has to recall that we still intend to find some
general principles valid in general irrespectively if the number $N$ of particles
is large or not. Unfortunately the standard connection between the two
pictures Schr\"odinger's and Heisenberg's fails down when $N$ is large in 
general. Remember, please, that when $N$ is large the gauge freedom
problem mentioned above goes into play: the two waves $\psi$ and 
$e^{i\xi(p)}\psi$ are to be considered equivalent. One can prove that instead
of normal Schr\"odinger picture with ordinary Schr\"odinger waves (Hilbert space 
vectors parametrically dependent on time) one has to deal with cross sections
of an appropriate Hilbert bundle $\mathcal{M}\triangle\mathcal{H}$  over 
whole spacetime $\mathcal{M}$. But, when using the 
Hilbert bundle $\mathcal{M}\triangle\mathcal{H}$ instead of
$\mathcal{H}$, the respective algebra of quantum operators acting in 
$\mathcal{M}\triangle\mathcal{H}$ has to be reducible, compare a subsequent
section where we give arguments for this fact. The spectrum of the center
of this algebra is just equal to the classical spacetime
$\mathcal{M}$. According to the above discussion 
this algebra corresponds to the noncommutative spacetime --- an object plying
the same role for quantum particles as the classical spacetime does for
classical bodies. 

In this way we have revealed the wave-particle duality
as a manifestation of the noncommutative structure of spacetime. The 
noncommutative algebra $\mathcal{A}$ corresponding to the noncommutative 
spacetime is obtained as the smallest von Neumann algebra
$\mathcal{A}^{CCR} \vee \mathcal{A}^{CCR}_{1}$ containing the von Neumann
canonical commutation algebra $\mathcal{A}^{CCR}$ of field operators and 
the appropriate maximally Abelian subalgebra $\mathcal{A}^{CCR}_{1}$ 
in the commutant of $\mathcal{A}^{CCR}$. The
center $\mathcal{A}^{CCR}_{1}$ (diagonal algebra) of $\mathcal{A}$ should 
not be trivial and corresponds to the classical spacetime.  

 Apparently, one can formulate a serious
objection against our conclusion that quantum particle lives in a 
spacetime with a noncommutative structure, with the structure closely
related to the algebra of quantum operators of the field corresponding
to the particle. Apparently one can say that each type of particle lives
in its own spacetime related to the corresponding type of field ---
which is a very strange idea rather. We maintain that the strange
effect of many coexisting spacetimes is apparent. Indeed. One should
note at this place that the representation of the algebra of
quantum field operators $\mathcal{A}^{CCR}$ we are interested in
should be reducible\footnote{Strictly speaking the matter is even more 
involved but we do not go into detail now in order to avoid excessive
mathematical complexities. We have assumed implicitly that the spacetime is 
compact so that the wave functions --- the respective cross sections of
the Hilbert bundle $\mathcal{M}\triangle\mathcal{H}$ --- do belong to the 
so-called direct integral Hilbert space $\int_{M} \mathcal{H}_{p} \, d\mu(p)$, 
$p\in\mathcal{M}$; compare the respective section of this Chapter. After this 
one can think of $\mathcal{A}^{CCR}$ and $\mathcal{A}$ as of acting in this 
direct integral Hilbert space.}. As such the algebra cannot act in the 
ordinary Fock space. The space is no longer a direct sum of symmetrized
(resp. anti-symmetrized) tensor products of one-particle Hilbert spaces,
in which the tensor product of $N$ factors occurs once and only once for
each natural $N$. This structure of Hilbert space is disturbed in our 
case. Let us recall that the symmetrization (resp. anti-symmetrization)
is deeply connected with the {\bf irreducibility} of the quantum algebra 
({\itshape H. Weyl, Gruppentheorie und Quantenmechanik, Leipzig,
Verlag von S. Hirzel, 1931}). In our case the representation is
{\bf reducible} so that the symmetrized and anti-symmetrized products may
appear within the same representation space! In the standard theory if the 
system is in symmetrized state it remains symmetrized forever, regardless of what 
influences may act upon it --- bosons and fermions do not mix each other.
In our case (in the limit of large $N$) the situation is substantially
different --- the symmetric and anti-symmetric states do mix each other. 
Therefore, the algebra $\mathcal{A}$ is much more universal object in comparison 
to any particular operator algebra of any specified kind of field in standard 
QFT. It, therefore, seems to be capable as to account for different kinds of 
particles. Moreover, it is an advantage that in the space of the 
representation of our algebra
the symmetrization (resp. anti-symmetrization) is disturbed. Remember, please, that
the symmetrization (resp. anti-symmetrization) reflects the the Bose-Einstein
(resp. Fermi-Dirac) correlations or the so-called entanglement of quantum states.
Thus, in our case the entanglement of states is disturbed which may correspond
to the fact that in practice for macroscopic bodies (when $N$ is large) the 
entanglement is negligible. It should be stressed that this fact is very difficult 
to explain within the ordinary QM and QFT.

Some other arguments concerning the apparent problem of many coexisting spacetimes 
corresponding to the various kinds of fields and particles seem advisable. 
The generalization of the wave-particle duality to a field-particle 
duality has dominated thinking in quantum theory 
for decades and has been heuristically useful in the development of 
elementary particle theory. Yet the belief in field-particle duality
as a general principle, the idea that to each particle there is a 
corresponding field and to each field a corresponding particle has also
been misleading. The number and the nature of different basic fields
is related to the charge structure, not to the empirical spectrum
of particles. For example, in the presently favored gauge theories
the fields are the carriers of charges like colour and flavour but 
are not directly associated to observed particles like electrons.         
The biunique field-particle correspondence is
therefore broken. Moreover, the spectacular issue of Connes and 
Lott ({\itshape Nucl. Phys. {\bf 18B}, 29, 1990}, see also {\itshape
A. Connes, loc. cit.}) that the so much a long list
of fields in the effective standard model can be considerably reduced 
with the cost of some noncommutative structure of spacetime involved.    
They considered the effective standard model, i.e. their 
analysis was confined to the classical context. The Lagrangean of this
effective theory is a combination of a five terms representing to 
independent contributions. Connes and Lott showed that 
this artificially complicated Lagrangean is a natural generalization
of the Maxwell-Dirac Lagrangean providing the appropriate noncommutative
structure of spacetime.      

Yet one has to be careful, however, and treat the specific structure of
$\mathcal{A}$ mentioned above as a prototype rather then as the ultimate
word one can say on this subject.

\vspace{1cm}

The fact that the concepts of ``spacetime coincidence'' and ``observation'' 
do require a thorough revision in QM was immediately noticed by the 
very founders of QM. It was especially evident in Bohr's writings.    
In the year 1925 he described the situation in the words: "From these 
results it seems to follow that, in the general problem of the quantum 
theory, one is faced not with a modification of the mechanical and 
electrodynamical theories describable in terms of the usual physical
concepts, but with an essential failure of the pictures in space and 
time on which the description of natural phenomena has hithero been 
based." (\emph{Nature}, {\bf 116}, p. 535.) Three years later he 
formulated the complementarity principle. As emphasized by Einstein
every observation or measurement ultimately rests on the 
coincidence of two independent events at the same spacetime point.
Now the quantum postulate implies that any observation
of atomic phenomena will involve an interaction with the agency of 
observation not to be neglected. On one hand -- Bohr stresses --
the definition of the state of a physical system, as ordinary 
understood, claims the elimination of all external disturbances. But 
in that case, according to the quantum postulate, any observation 
will be impossible, and, above all, the concepts of space and time 
lose their immediate sense. On the other hand -- he concludes -- if 
in order to make observation possible we permit certain interactions 
with suitable agencies of measurement, not belonging to the system, 
an unambiguous definition of the state of the system is naturally no 
longer possible, and there can be no question of causality in the 
ordinary sense of the word (Heisenberg uncertainty principle). 
Concluding, we have the complementarity alternative: \emph{Either} 
quantum particle is describable in terms of space and time but the state 
of particle is not well defined \emph{or} the state of quantum 
particle is well defined but its description in terms of space and 
time impossible. We propose to interpret this complementarity 
principle as indicating that the spacetime event, in the classical
sense of the word, cannot be ascribed to quantum particle according
to our analysis presented above. The quantum particle lives in
a noncommutative spacetime. The event in (classical) spacetime can 
be ascribed to a body which we would like to call classical -- 
made of many quantum particles.
    
\section{Gauge Symmetry}

\subsection{Gauge Freedom and Heisenberg Commutation Relations}

Let us back to the ordinary QM applied to a system with finite number
of degrees of freedom. To be consequent we have to restrict our
consideration to non-relativistic case for the moment --- please, 
remember that any relativistic quantum system has to constitute 
a quantum field system with infinite number of degrees of freedom. 
The quantum states of our system constitute a Hilbert space 
$\mathcal{H}$. The inner product $(\centerdot,\centerdot)$ of the 
space is an relevant part of structure plying important role
in the physical interpretation. Namely, the only contribution 
of the the vector $\varphi$ of the Hilbert space $\mathcal{H}$ to 
any measurable effects is inscribed into the set of numbers
\begin{displaymath}
\frac{\vert(\phi, \varphi)\vert^{2}}{(\phi,\phi)(\varphi, \varphi)},
\, \, \,  \phi \in \mathcal{H}.   
\end{displaymath}
Suppose we have an ideal source which prepares an ensemble
in apure state, described by an element of $\mathcal{H}$.
In particular if in addition we have an ideal detector, giving a 
yes-answer in a pure state $\varphi$ and the answer no in the orthogonal 
complement to $\phi$, then the probability of detecting an event
in this set up of source and detector is given by        
\begin{displaymath}
\frac{\vert(\phi, \varphi)\vert^{2}}{(\phi,\phi)(\varphi, \varphi)}.   
\end{displaymath}  
Therefore, two vectors $\varphi$ and $e^{i\xi}\varphi$ differing 
by a mere constant phase are indistinguishable. This is very important 
for the whole structure of quantum theory. For example, any kind of 
algebra of some quantities, like the algebra of classical observables 
(functions $f(p,q)$ on the phase space with ordinary 
point wise operations) --- which is commutative ---
will not necessary be commutative if unitary represented as operator algebra
acting in the Hilbert space of states. Let us explain how it follows
from the constant-phase-equivalence of the two vectors $\varphi$
and $e^{i\xi}\varphi$, $\xi$ being a constant real number. Consider
the set of invertible elements of the algebra in question, which 
constitutes a commutative group in our case. Because of the 
constant-phase-equivalence the two operators $A$ and $e^{i\xi}A$ are
equivalent giving the same average values and having the same spectra.
Therefore in general the relation 
\begin{displaymath}
AB = e^{i\xi(A,B)}BA
\end{displaymath}    
holds for representations acting in $\mathcal{H}$ instead of
\begin{displaymath}
AB = BA.
\end{displaymath} 
Such a representation is mostly called {\itshape ray representation}.
Let us consider representation $U_{i}$ and $V_{i}$ of the $2n$-dimensional
Abelian group of phase coordinates $(p_{1}, \ldots p_{n},q_{1}, 
\ldots q_{n})$ in the classical phase space. The algebra of classical phase 
coordinates is generated by the one-parameter groups corresponding to
$p_{i}$ and $q_{i}$ which should hold also for the quantum representation. 
If one assumes the unitary representation
to be {\bf irreducible} then one gets the relations 
\begin{eqnarray}
U_{i}V_{k} = & e^{i\xi(p_{i},q_{k})}V_{k}U_{i}, \nonumber \\
U_{i}U_{k} = & e^{i\xi(p_{i},p_{k})}U_{k}U_{i}, \nonumber \\
V_{i}V_{k} = & e^{i\xi(q_{i},q_{k})}V_{k}V_{i}, \nonumber 
\end{eqnarray}  
equivalent to the Heisenberg commutation relations and moreover the 
representation is unique up to unitary equivalence. That is, denoting the 
generators of $U_{i}$ and $V_{k}$ by $P_{i}$ and $Q_{k}$ we get from the above
relations the following result 
\begin{eqnarray}
Q_{i}P_{k} - P_{k}Q_{i} = & i\delta_{ik}, \nonumber \\
Q_{i}Q_{k} - Q_{k}Q_{i} = & 0, \nonumber \\
P_{i}P_{k} - P_{k}P_{i} = & 0, \nonumber 
\end{eqnarray}
i. e. the Heisenberg commutation relations. This is a well known result
noticed by Weyl ({\itshape loc. cit.}) investigated further by v. Neumann, Rellich and
Stone. The part of the statement concerning the uniqueness is mostly called 
von Neumann's uniqueness theorem. 
 
From the mathematical point of view this Weyl's result is nothing else but
a special case of Bargmann's theory of ray representations ({\itshape loc. 
cit.}) applied to the Abelian group of canonical coordinates (i.e. to the 
translation group in the phase space). 

The requirement that the set of $2n$ operators $P_{1} \ldots P_{n},Q_{1}, 
\ldots ,Q_{n}$ should be irreducible is very important in the proof of the above
statement of Weyl but also in the whole of QM laws! This postulate is to 
be added to the Heisenberg commutation rules as an essential supplement.
For example in QM there is a standard rule for description of systems composed
of several individual parts (i. e. subsystems). Suppose a system $A$
to be composed of two parts $B$ and $C$. The states of the system and the 
subsystems are to be represented by vectors in the Hilbert spaces 
$\mathcal{H}_{A}$, $\mathcal{H}_{B}$ and $\mathcal{H}_{C}$ respectively. 
Then, the general rule says first that $\mathcal{H}_{A} = 
\mathcal{H}_{B}\otimes\mathcal{H}_{C}$ is the ordinary tensor product
of $\mathcal{H}_{B}$ and $\mathcal{H}_{C}$. Second, in accord to the rule,
the only hermitian operators which have physical significance depend 
symmetrically on the two subsystems. Let us consider the very special
situation of this kind when both $A$ and $B$ are quantum particles 
of the same kind. We, thus, have 12 matrices $P_{1} \ldots P_{6},Q_{1}, 
\ldots ,Q_{6}$ fulfilling Heisenberg commutation rules. In general
they are reducible in accordance with the decomposition 
\begin{displaymath}
\mathcal{H}_{A} =
\mathcal{H}_{B}\otimes\mathcal{H}_{B} = \mathcal{H}_{B}\otimes_{S}\mathcal{H}_{B}
+ \mathcal{H}_{B}\otimes_{A}\mathcal{H}_{B}
\end{displaymath} 
of the Hilbert space of the system $A$ into the symmetric and anti-symmetric 
tensor product of the Hilbert spaces corresponding to the particles $B$ and
$C=B$ (remember, please, that all $P_{i}$ and $Q_{k}$ depend symmetrically
on subsystems $B$ and $C=B$). Experimental evidence tells us that there are
only two kind of particles those with states either in 
$\mathcal{H}_{B}\otimes_{S}\mathcal{H}_{B}$ or 
$\mathcal{H}_{B}\otimes_{A}\mathcal{H}_{B}$. Thus the reducibility of the set
of operators $P_{1} \ldots P_{n},Q_{1}, \ldots ,Q_{n}$  would contradict 
experiment. 

Having given this comment on irreducibility let us back
to the constant-phase equivalence and the Weyl's scheme within which
we have obtained the Heisenberg commutation rules from this equivalence.
It is remarkable that the quantization of the problem of several 
particles also falls within this general scheme even for fermions. 
In dealing with it we are interested in that Abelian group whose
basic elements $p_{\alpha}, q_{\alpha}$ are all of order 2 in the 
fermionic case. Such a group consists of the totality of monomials
\begin{displaymath}
p_{1}^{n_{1}}q_{1}^{n_{2}}p_{2}^{n_{3}}q_{2}^{n_{4}}\centerdot \ldots,
\end{displaymath}
where $n_{k} = 1$ or 0 and $n_{1} + n_{2} + \ldots \leq N$ (total number of 
particles), compare ({\itshape Weyl, loc. cit.}). In virtue that 
the two representations $P_{\alpha}, Q_{\alpha}$ and 
$P'_{\alpha} =  e^{\beta(p_{\alpha})}P_{\alpha}, Q'_{\alpha} 
= e^{\beta(q_{\alpha})}Q_{\alpha}$ are equivalent the gauge (exponent $\beta$) 
can be so chosen that the corresponding operators of the irreducible
ray representation satisfy the anti-commutation rules          
\begin{eqnarray}
Q_{i}P_{k} + P_{k}Q_{i} = & i\delta_{ik}, \nonumber \\
Q_{i}Q_{k} + Q_{k}Q_{i} = & 0, \nonumber \\
P_{i}P_{k} + P_{k}P_{i} = & 0. \nonumber 
\end{eqnarray}    

In passing to the relativistic QM we have to account for systems with
infinite degrees of freedom. To obtain the Heisenberg rules within
the same general scheme presents a very sophisticated mathematical
problem. In the case of a free field, however, it is quite possible.
But now the von Neumann uniqueness theorem no longer holds. There 
is a vast of inequivalent irreducible representations of Heisenberg 
commutation rules. We have to impose additional condition that the 
representation possesses a cyclic vector with some peculiar properties,
namely the vacuum.   

Because of this results one can see that the constant-phase equivalence
of Hilbert space vectors lies at the very heart of QM. On the other 
hand there is a very natural way of generalizing this Weyl's scheme.
For example in the non-relativistic QM we can use the Schr\"odinger 
picture. But then the two wave functions $\psi$ and $e^{i\xi(t)}\psi$
are equivalent. Of course one has to assume that the Schr\"odinger
equation is endowed with the appropriate time dependent gauge freedom,
but this is very realistic and one cannot exclude such situation,
compare the next subsection. For example when considering wave equation
in gravitational field (of course non-relativistic field in this case)
the time dependent gauge freedom is unavoidable.        
Because the constant-phase equivalence is so important one cannot
ignore it and the consequences of this generalization should be
investigated. Even more, one can go still a step further if starting
from QFT instead of QM, as we argued in the first section of this 
Chapter.                

It is,thus, justified to think of this phase-equivalence as of a
special kind of gauge symmetry. In this way the Weyl's constant-phase
equivalence is a special kind of gauge symmetry, namely the 
constant-phase symmetry ---the simplest possible one.

\subsection{Time Dependent Gauge Freedom}\label{Time}

In this subsection we carry out general analysis of the representation 
$T_{r}$ of a covariance group, and compare it with the representation of a 
symmetry group. We also describe correspondence between the space of wave 
functions $\psi(\vec{x},t)$ and the Hilbert space. Here the analysis is 
performed in the non-relativistic case. 

Before we give the general description, it will be instructive to investigate 
the problem for a free particle in the flat Galilean spacetime. The set of 
solutions $\psi$ of the Schr\"odinger equation, which are admissible in 
Quantum Mechanics, is precisely given by 
\begin{displaymath}
\psi(\vec x,t)=(2\pi)^{-3/2} \int \varphi(\vec k)e^{-i\frac{t}{2m}\vec{k}
\centerdot \vec k+i\vec k\centerdot\vec x} \, {\ud}^{3}{k},
\end{displaymath}
where $p=\hslash k$ is linear momentum and $\varphi(\vec{k})$ is any square 
integrable function. The functions $\varphi$ (wave functions in the "Heisenberg 
picture") form a Hilbert space $\mathcal{H}$ with the inner product
\begin{displaymath}
(\varphi_{1}, \varphi_{2})=\int \varphi_{1}^{*}(\vec{k})\varphi_{2}(\vec{k}) \, 
{\ud}^{3}{k}.
\end{displaymath}
The correspondence between $\psi$ and $\varphi$ is one-to-one. 

In general, however, the construction fails if the Schr\"odinger equation possesses 
nontrivial gauge freedom. Let us explain it. For example, the above construction 
fails for the non-relativistic quantum particle in the curved Newton-Cartan 
spacetime. Besides, in this spacetime we do not have any plane wave, see 
({\itshape J. Wawrzycki, Int. Jour. of Theor. Phys. {\bf 40}, 1595 (2001)}). 
Thus, there does not exist any natural counterpart for the Fourier 
transform. However, we do not need to use the Fourier transform. What is the role 
of the Schr\"odinger equation in the above construction of $\mathcal{H}$?
Please note that in general 
\begin{displaymath}
\Vert \psi \Vert^{2} \equiv
\int \psi^{*}(\vec{x},0)\psi(\vec{x},0) \, {\ud}^{3}{x} = (\varphi,\varphi) 
\end{displaymath}

\begin{displaymath}
=\int \psi^{*}(\vec{x},t)\psi(\vec{x},t) \, {\ud}^{3}{x}.
\end{displaymath}
This is in accordance with the Born interpretation of $\psi$. Namely, if 
$\psi^{*}\psi(\vec{x},t)$ is the probability density, then
\begin{displaymath}
\int \psi^{*}\psi \, {\ud}^{3}{x}
\end{displaymath}
has to be preserved over time. In the above construction, the Hilbert space 
$\mathcal{H}$ is isomorphic to the space of square integrable functions 
$\varphi(\vec{x})\equiv \psi(\vec{x},0)$, namely the set of square integrable
initial data for the Schr\"odinger equation, cf. e.g. 
({\itshape D. Giulini, {\sl States, Symmetries and Superselection}, 
in: {\sl Decoherence: Theoretical, Experimental 
and Conceptual Problems}, (Lecture Notes in Physics, Springer Verlag 2000), 
page 87.}). 
The connection between $\psi$ and $\varphi$ is given by the time evolution 
operator $U(0,t)$ (equivalently by the Schr\"odinger equation):
\begin{displaymath}
U(0,t)\varphi=\psi.
\end{displaymath}
The correspondence between $\varphi$ and $\psi$ has all formal properties, such 
as in the Fourier construction above. Of course, the initial data for the 
Schr\"odinger equation do not cover the whole Hilbert space 
$\mathcal{H}$ of square integrable functions, but the time evolution given by 
the Schr\"odinger equation can be uniquely extended over the whole Hilbert space 
$\mathcal{H}$ by the unitary evolution operator $U$. 

The construction can be applied to the particle in the Newton-Cartan spacetime. 
As we implicitly assumed, the wave equation is such that the set of its 
admissible initial data is dense in the space of square integrable functions
(we need this for the uniqueness of the extension). Because of the Born 
interpretation, the integral
\begin{displaymath}
\int \psi^{*}\psi \, {\ud}^{3}{x}
\end{displaymath}
has to be preserved over time. Let us denote the space of the square-integrable 
initial data $\varphi$ on the simultaneity hyperplane $t(\vec{x},t)=t$ by 
${\mathcal{H}}_{t}$. 
Then, the evolution is an isometry between ${\mathcal{H}}_{0}$ and 
${\mathcal{H}}_{t}$. But such an isometry has to be a unitary operator, 
and the construction is well defined, \emph{i.e.} the inner product of two 
states corresponding to the wave functions $\psi_{1}$ and $\psi_{2}$
does not depend on the choice of ${\mathcal{H}}_{t}$. Let us mention that 
the wave equation has to be linear in accordance with the Born interpretation 
of $\psi$ (since any unitary operator is linear the time evolution operator is 
linear as well). 
The space of wave functions $\psi(\vec{x},t) = U(0,t)\varphi(\vec{x})$
isomorphic to the Hilbert space ${\mathcal{H}}_{0}$ of $\varphi$'s is commonly
called the "Schr\"odinger picture".  

However in general, the connection between $\varphi(\vec{x})$ and  
$\psi(\vec{x},t)$ is not unique if the wave equation possesses a gauge freedom. 
Namely, let us consider two states $\varphi_{1}$ and $\varphi_{2}$ and ask 
when these two states are equivalent, and indistinguishable. The answer is 
that they are equivalent if 
\begin{displaymath}
\vert(\varphi_{1},\varphi)\vert \equiv \Big\vert\int \psi_{1}^{*}(\vec{x},t)
\psi(\vec{x},t) \, {\ud}^{3}{x}\Big\vert = \vert(\varphi_{2},\varphi)\vert 
\end{displaymath}

\begin{equation}\label{row}
\equiv \Big\vert\int \psi_{2}^{*}(\vec{x},t)\psi(\vec{x},t) \, 
{\ud}^{3}{x}\Big\vert,
\end{equation}
for any state $\varphi$ from $\mathcal{H}$, or for all $\psi=U\varphi$ 
($\psi_{i}$ are defined to be = $U(0,t)\varphi_{i}$). By substituting $\varphi_{1}$ 
and then $\varphi_{2}$ for $\varphi$ and making use of the Schwarz's inequality, 
one gets: $\varphi_{2}=e^{i\alpha}\varphi_{1}$, where $\alpha$ is any 
constant\footnote{This gives the conception of the ray, introduced to Quantum 
Mechanics by Hermann Weyl
[H. Weyl, {\sl Gruppentheorie und Quantenmechanik}, 
Verlag von S. Hirzel in Leipzig (1928)]: a physical 
state does not correspond uniquely to a normed state $\varphi \in \mathcal{H}$, 
but it is uniquely described by a \emph{ray}; two states belong to the same 
ray if they differ by a constant phase factor.}.
The situation for $\psi_{1}$ and $\psi_{2}$ is however different. 
In general, condition (\ref{row}) is fulfilled if 
\begin{displaymath}
\psi_{2}=e^{i\xi(t)}\psi_{1}
\end{displaymath}
and the phase factor can depend on time. Of course, this has to be consistent with 
the wave equation, that is, together with a solution $\psi$ of the wave equation, 
the wave function $e^{i\xi(t)}\psi$ is also a solution of the appropriately 
gauged wave equation. \emph{A priori} one cannot exclude the existence of such 
a consistent time evolution. This is not a new observation, as it was 
noticed by John von Neumann\footnote{J. v. Neumann, {\sl Mathematical Principles 
of Quantum Mechanics}, University Press, Princeton (1955). He did not mention 
gauge freedom on that occasion. However, gauge freedom 
is necessary for the equivalence of $\psi_{1}$ and $\psi_{2} 
= e^{i\xi(t)}\psi_{1}$.}, but it seems that it has never 
been deeply investigated (probably because the ordinary non-relativistic 
Schr\"odinger equation has gauge symmetry with constant $\xi$). 
The space of waves $\psi$ describing the system cannot be reduced in the above 
way to any fixed Hilbert space $\mathcal{H}_{t}$ with a fixed $t$. So, 
the existence of the nontrivial gauge freedom leads to the following

\vspace{1ex}

{\bf Hypothesis}. \emph{The two waves} $\psi$ \emph{and} $e^{i\xi(t)}\psi$ 
\emph{are quantum-mechanically indistinguishable}.

\vspace{1ex}

Moreover, we are obliged to use the whole Hilbert bundle $\mathcal{R}\triangle 
\mathcal{H}: t \to \mathcal{H}_{t}$ over the time instead of a fixed Hilbert
space $\mathcal{H}_{t}$, with the appropriate cross-sections $\psi$ as the waves
(see the next section for details). 

Let us consider now an action $T_{r}$ of a group $G$ in the space of waves $\psi$.    
Before we infer some consequences of the assumption that $G$ is a symmetry 
group, we need to state a:

\vspace{1ex}

{\bf Classical-like  postulate}. \emph{Group $G$ is a symmetry 
group if and only if the wave equation is invariant under the transformation
$x' = rx, r\in G$ of independent variables and the transformation $\psi'= 
T_{r}\psi$ of the wave function}.

\vspace{1ex}
 
The above postulate is indeed commonly accepted in Quantum Mechanics even when
the gauge freedom is not excluded. But it is a mere application of the symmetry
definition for a classical field equation applied to the wave equation
without any change. The wave $\psi$ is not a classical quantity, such as e.g.
electromagnetic intensity. The above {\bf Hypothesis} is not
true for classical fields, and we have to be careful in forming the appropriate
postulate for the wave equation compatible with the {\bf Hypothesis}.   
Namely, the two wave equations differing by a mere gauge are indistinguishable.
We call them \emph{gauge-equivalent}. It is therefore natural to assume the

\vspace{1ex}

{\bf Quantum postulate}. \emph{Group $G$ is a symmetry 
group if and only if  the transformation $x' = rx, r\in G$ of independent 
variables and the transformation $\psi'= T_{r}\psi$ of the wave function 
transform the wave equation into a gauge-equivalent one}.

\vspace{1ex}

Please note that not all possibilities admitted by the {\bf Hypothesis}
are included in the {\bf Classical-like postulate}. 

From the {\bf Classical-like postulate} it follows that $\psi$
as well as $T_{r}\psi$ are solutions of exactly the same wave equation,
in view of the invariance of the equation. Therefore, $\psi$ and 
$T_{r}\psi$ belong to the same "Schr\"odinger picture", so that
\begin{displaymath}
T_{r}T_{s}\psi = e^{i\xi(r,s)}T_{rs}\psi,
\end{displaymath}
with $\xi = \xi(r,s)$ independent of time $t$! This is in accordance
with the known theorem that

\vspace{1ex}

\begin{twr} If $G$ is a symmetry group, then the phase factor $\xi$ should 
be time-independent. \end{twr}

\vspace{1ex}

But if we start from the {\bf Quantum postulate}, we obtain 
instead
\begin{equation}\label{rowt}
T_{r}T_{s}\psi=e^{i\xi(r,s,t)}T_{rs}\psi
\end{equation}
and get 

\vspace{1ex}

{\bf Theorem 1'} \, \emph{If $G$ is a symmetry group, then the phase 
factor $\xi = \xi(r,s,t)$ is time-dependent in general}. 

\vspace{1ex}

In this paper we propose to accept the {\bf Quantum postulate}, which is 
compatible with the {\bf Hypothesis}, and is more in spirit of Quantum 
Mechanics than the {\bf Classical-like postulate}. It should be noted 
that in the special case when  gauge freedom degenerates to
the case of constant phase, the {\bf Quantum postulate} is equivalent to
to the {\bf Classical-like postulate}. 

We shall resolve the following paradox.
Namely, a natural question arises why the phase factor $e^{i\xi}$ in 
(\ref{rowt}) is time-independent for the Galilean group (even when
the Galilean group is considered as a covariance group). The explanation of the 
paradox is as follows. The Galilean covariance group $G$ induces the 
representation
$T_{r}$ in the space $\mathcal{R}\triangle \mathcal{H}$ 
and fulfills (\ref{rowt}). But, as we will show later on, 
the structure of $G$ is such that 
there always exists a function $\zeta(r,t)$ continuous in $r$ and 
differentiable in $t$, with the help of which one can define a new equivalent 
representation $T'_{r}=e^{i\zeta(r,t)}T_{r}$ fulfilling
\begin{displaymath}
T'_{r}T'_{s}=e^{i\xi(r,s)}T'_{rs} 
\end{displaymath}
with a time-independent $\xi$. The representations $T_{r}$ and $T'_{r}$ 
are equivalent because $T'_{r}\psi$ and $T_{r}\psi$ are equivalent for all 
$r$ and $\psi$. However, this is not the case in general, when the exponent 
$\xi$ depends on time, and this time dependence cannot be eliminated in 
the same way as for the Galilean group. We have a similar situation when we try 
to find the most general wave equation for a non-relativistic quantum particle 
in the Newton-Cartan spacetime. The relevant covariance group in this case
is the Milne group which possesses representations with time-dependent $\xi$ not 
equivalent to any representations with a time-independent $\xi$. Moreover, 
the only physical representations of the Milne group are those with 
time-dependent $\xi$.

\subsection{Spacetime Dependent Gauge Freedom}

There is a physical motivation to investigate 
representations $T_{r}$ fulfilling (\ref{rowt}) with $\xi$ depending on 
spacetime point $p$:
\begin{equation}\label{rowX}
T_{r}T_{s}=e^{i\xi(r,s,p)}T_{rs}.
\end{equation}
We have sketched the motivation in the first section.
We have argued there, that the two wave functions $\psi$ and $\psi' = 
e^{i\xi(p)}\psi$ are indistinguishable in the sense that 
they give 
the same transition probabilities: $\vert(\psi,\phi)\vert^{2} = 
\vert(\psi',\phi)\vert^{2}$ for any $\phi$. One should provide,
however, that we are sufficiently fare away from the vacuum.
This additional assumption is an immediate consequence of the 
structure of states in the Fock space as well the form of the inner
product in that space:
\begin{displaymath}
\psi = \left( \begin{array}{ll}
c & \\
\psi_{1}(x) & \\
\psi_{2}(x_{1},x_{2}) & \\
\ldots &
\end{array} \right)
\end{displaymath}
and
\begin{displaymath}
(\psi,\psi') = c^{*}c' + \int_{\mathbb{R}^{3}}\psi_{1}^{*}\psi_{1}' \, d^{3}x  + 
\int_{\mathbb{R}^{2\times 3}}\psi_{2}^{*}\psi_{2}' \, d^{3}x_{1}d^{3}x_{2} + \ldots
\end{displaymath}   
in which the argument $x$ of the one-particle state $\psi_{1}$ contains 
the ordinary space coordinates. If one uses the Schr\"odinger picture
the spacetime-dependent-phase equivalence effect is of primary
importance, as one can expect by comparison with the previous subsection.
Probably it would be superfluous to present in detail that in this
case one is forced to use the whole Hilbert bundle 
$\mathcal{M}\triangle \mathcal{H}$ over spacetime $\mathcal{M}$ and 
respective cross-sections as the wave functions $\psi$ (see the next 
section for definitions). We do not present details as the reasoning
is a simple analogue of that performed in the previous subsection.
 
Rather we concentrate on the heart of the whole problem, that is,
on the spacetime-dependent-phase equivalence which seems more advisable.
As we have said application of that equivalence is justified if working
far from the vacuum state, when there is quite a number of particles 
present. But then the  justification of this equivalence principle 
is the same as that of the fact that spacetime coordinates are 
$c$-numbers commuting with ``everything'', so that
the greater the number of particles the more commuting are the
spacetime coordinates.  This is natural and agrees with the well
established knowledge that when dealing with one particle (within QM)
the spacetime coordinates are not mere parameters and do not commute
with ``everything''; but, in passing to quantum field by canonical
quantization (appropriate for many particles) the spacetime coordinates
are ordinary $c$-numbers commuting with each other and all other 
quantities. Yet the state of affairs is not quite satisfactory and
there is a problem which calls for a further analysis. As we have 
argued in first section the QFT is expected to work perfectly 
along with QM laws applied to infinite number of degrees of
freedom when there is a very few particles present, i. e. near the 
vacuum state. This is the case in practice, for example, when computing
both the Lamb shift and the anomalous magnetic moment --- when we are 
dealing with one-electron problem. It seems, therefore, that application of our
equivalence is not justified, but at the same time the commutativity
of spacetime coordinates or their $c$-number character is not
justified too! This contradict the canonical field quantization rule
in which the coordinates do form a $c$-numbers! In this way
we arrive at the puzzle of spacetime coordinate status which we 
shall try to resolve now.

First of all we should note that the spacetime-coordinates-problem does 
not exist when the quantum field is free --- compare the first section. 
It goes into play when interaction is taken into account. We confine 
ourselves to QED in order to be more specific. One should like to work
within the ordinary QM perturbation theory considered as causing 
transitions between the stationary states say of the free field. 
The ordinary QED in Schr\"odinger picture, however, presents so much a 
departure from logic in applying the QM perturbation theory, that it is 
even impossible to work within this formulation of QED. In general
when a realistic interaction is present, so violent in the high 
frequencies, the ordinary picture of perturbation as causing transitions
between ordinary stationary states of the free field (as in the anomalous 
magnetic moment) is destroyed and does require some extra caution.
The Sch\"odinder picture is unsuited for dealing with
QED, because the vacuum fluctuations play such a dominant role in it.
Still we shall try to reformulate the QED so as to be as much compatible
with ordinary QM laws as possible. Such a reformulation was proposed by
Dirac in his excellent book ({\itshape {\sl Lectures of Quantum Field Theory},
Academic Press, New York, 1966}; see also the last Chap. of the Fourth 
revised 1981 ed. of {\itshape The Principles of Quantum Mechanics}).
We have no room here to present the reformulation but we should quote some 
Dirac's statements at least, which are of importance in our discussion. 
Suppose the ket $|Q\rangle$ represents a state for which there are no photons,
electrons, or positrons present. One would be inclined to suppose this state 
to be the perfect vacuum, but it cannot be, because it is not stationary.
For it to be stationary we should need to have 
\begin{displaymath}
H|Q\rangle = \lambda |Q\rangle
\end{displaymath}  
with $\lambda$ a number and $H$ the Hamiltonian of QED. Now $H$ contains 
the terms (we use the standard notation)
\begin{equation}\label{terms}
-e \int \bar{\psi}\alpha_{r}\mathcal{A}^{r}\psi \, d^{3}x
+\frac{1}{2} \int\!\!\!\int \frac{j_{0x}j_{0x'}}{|x-x'|} \, d^{3}xd^{3}x', 
\end{equation}      
which do not give numerical factors when applied to $|Q\rangle$. If we start
with the no-particle state it does not remain the no-particle state.
Particles get created where none previously existed, their energy 
coming from the interaction part of the Hamiltonian. Let us call the no-particle
state at a certain time by $|Q\rangle$. In order to study this spontaneous 
creation of particles, one takes the ket $|Q\rangle$ as initial in the
Schr\"odinger picture and treat the terms (\ref{terms}) as a perturbation
giving rise to a probability of the state $|Q\rangle$  jumping into 
another state, in accordance with the ordinary perturbation theory of QM.
The first term resolved into its Fourier components --- the photon, electron
and positron creation operators --- contains a part   
\begin{equation}\label{term1}
-e(\alpha_{r})^{ab} \int\!\!\!\int \mathcal{A}^{r}_{k} \bar{\xi}_{ap}
\zeta_{b p+k\hslash} \, d^{3}kd^{3}p, 
\end{equation}       
causing transitions and corresponding to emission of a photon (creation 
operator $\mathcal{A}^{r}_{k}$) and simultaneously to creation of an 
electron-positron pair (creation operators $\bar{\xi}_{ap}$ and 
$\zeta_{b p+k\hslash}$). After a short time the transition probability
is proportional to the squared length of the ket formed by multiplying
(\ref{term1}) into the initial ket $|Q\rangle$. But this length is 
infinite, so the transition probability is infinite. The second term
of (\ref{terms}) contains contributions with two electron-positron 
pairs created simultaneously. Again the transition probability due to
this term is also infinite. One can conclude that the state $|Q\rangle$
is is not even approximately stationary. Even with a cutoff the no-particle 
state $|Q\rangle$ is not approximately stationary. This is why the above
procedure presents so much a departure from the ordinary perturbation
theory of QM, and seems to be not logically justified. Dirac proposes
another way of dealing with QED, which is a less of departure from 
ordinary QM. From the the above calculations --- says Dirac --- it
follows that the no-particle state $|Q\rangle$ differs very much from 
the vacuum state. The ``vacuum'' state must contain many particles,
which may be pictured as a state of transient existence with violent
fluctuations. Let us introduce the ket $|V\rangle$ to represent
the ``vacuum'' state. It is the eigenket of $H$ belonging to the
lowest eigenvalue. Here and subsequently $H$ denotes the Hamiltonian 
modified by the cutoff. One might try to calculate $|V\rangle$ as a 
perturbation of of the ket $|Q\rangle$, but such a method would be of 
doubtful validity, because the difference between $|V\rangle$ and 
$|Q\rangle$ is not small. No satisfactory way of calculating $|V\rangle$
is known. In any case the result would depend strongly on the cutoff,
and since the cutoff is unspecified the result would not be a definite 
one. It follows that we must develop the theory without knowing 
$|V\rangle$. This is not a great hardship --- argues further Dirac ---
because we are not manly interested in the ``vacuum''. We are mainly 
interested in states which differ from the ``vacuum'' $|V\rangle$
through having a few particles present in addition to those associated
with the vacuum fluctuations, and we want to know how this extra
particles behave. For this purpose we focus our attention on an operator
$K$ representing the creation of the extra particles, so that the state
we are interested in appears $K|V\rangle$. We do not now how the ket 
$|V\rangle$ varies with time in the Schr\"odinger picture, since we do 
not now the lowest eigenvalue of $H$. To avoid this difficulty we work
in the Heisenberg picture in which $|V\rangle$ is constant. We then 
require $K|V\rangle$ to represent another state in the Heisenberg 
picture and thus to be another constant ket. This leads to
\begin{equation}\label{evolution}
\frac{dK}{dt} \,\,\, \textrm{or} \,\,\, 
i\hslash \frac{\partial K}{\partial t} + KH - HK =0.
\end{equation}
We now have each physical state determined by a solution $K$ of
(\ref{evolution}). Dirac ({\itshape loc. cit.}) proceeded along 
these lines and built a theory more compatible with the standard QM.
He was able to calculate the Lamb shift as well as the anomalous 
magnetic moment within this theory. Thus it is an open question if
we are close to the ``true'' vacuum $|0\rangle$ when evaluating the 
anomalous magnetic moment. From the above calculations we expect rather
that the state $|V\rangle$ is considerably far removed from the 
true vacuum $|0\rangle$ (no-particles present). But if we are 
sufficiently far from the vacuum to ensure the commutativity of sapacetime
coordinates (as in the canonical field quantization) we will at the 
same time ensure the justification for our 
spacetime-dependent-equivalence of states! Is it therefore possible
that we should use the Hilbert bundle 
$\mathcal{M}\triangle\mathcal{H}$ with appropriate cross sections 
instead of ordinary Hilbert space $\mathcal{H}$ with vectors of 
$\mathcal{H}$? It depends if the difference between $|V\rangle$ 
and the true vacuum is large or small --- the fact on which we
may speculate only. Anyway one has the following alternative:
\begin{displaymath}
\begin{array}{rccc}
      & \textrm{{\scriptsize Small difference between $|0\rangle$ and $|V\rangle$}}      &              & \textrm{{\scriptsize Large difference between $|0\rangle$ and $|V\rangle$}} \\
\textrm{{\bf Either}}   & \textrm{\scriptsize $\mathcal{H}$}  &  \textrm{{\bf or}} & \textrm{\scriptsize $\mathcal{M}\triangle\mathcal{H}$} \\
 & \textrm{{\scriptsize noncommuting spacetime coordinates}} &               & \textrm{\scriptsize commuting spacetime coordinates}.
\end{array}
\end{displaymath}
On the left hand side we have a case in which the spacetime-dependent
equivalence is not always justified. In that case one actually works
near the true vacuum $|0\rangle$ when considering one-particle problems
like the anomalous magnetic moment. Ordinary QM with states as vectors 
(or Weyl's rays) in ordinary Hilbert space is justified then, 
but the spacetime coordinates are not a mere $c$-numbers in it.
On the right hand side we have a theory in which the space-time-dependent
equivalence is in general justified. States are the appropriate cross sections 
$\psi: \, \mathcal{M}\ni p \to \psi_{p} \in \mathcal{H}_{p}$ in a Hilbert bundle 
$\mathcal{M}\triangle\mathcal{H}: \,\mathcal{M} \ni p \to \mathcal{H}_{p}$ 
over spacetime in which the respective Hilbert spaces $\mathcal{H}_{p}$ 
play a role rather.                    
Now, one can see what a peculiar object is the QED. Namely, it is partly
on the left hand side of the alternative as it use the basic methods of 
ordinary QM with $\mathcal{H}$ and at the same time it is at the right 
hand side of the alternative as a canonically
quantized version of the classical field with {\bf commuting spacetime 
coordinates}. Strictly speaking there are some troubles in constructing ordinary
Hilbert space $\mathcal{H}$ for QED, as was pointed by Dirac ({\itshape loc. cit.})
even within his method of treatment mentioned above. Moreover, in the canonical
quantization one is trying at the outset to implement the ordinary QM to a system
with infinite number of degrees of freedom. We believe that the problem just 
mentioned mirrors an important physical truth. {\bf It strongly suggest that
the spacetime coordinates in QED should not be perfectly commuting quantities,
and that the less commuting they are the stronger is the cutoff effect}. Indeed,
if they are noncommuting, then we are on the left hand side of the alternative,
where the difference between $|0\rangle$ and $|V\rangle$ is small. But this is 
possible only if the cutoff is large. On the other hand if they are commuting,
then  we are on the right hand side of the alternative where the difference
between $|0\rangle$ and $|V\rangle$ is large, i. e. when the cutoff is small 
and not so violent. As is well known, the infinities in QED originate from the 
fact that we pass to the limit 
zero for the space and time intervals involved. But if the spacetime 
coordinates were noncommuting then the structure of spacetime would be more 
elaborate and the limit process would be meaning less. {\bf This suggest that the
cutoff process reflects some important physical phenomena}.\footnote{People usually
think that the cutoff is a man-made process which one has to perform
due to the fact that the theory at our disposal is incapable of the high energy 
processes. This opinion was justified in the early seventies, when the strong
interactions were out of our scope. One could hope that the strong interactions
will cancel the infinities. But now it seems that this hope turned out to be
vain. Strong interactions do not cancel the infinities. Paradoxically QED
serves now as the paradigm for a successful quantum field theory.}          

It is therefore advisable to construct a theory which like QED lies somewhat 
on the both sides of the above alternative. Namely, it this theory the states
as cross sections (looking at the right hand side) should compose at the same
time a Hilbert space (looking at the left hand side). There is a natural 
Hilbert space composed of appropriate
cross sections of a Hilbert bundle in mathematics, namely the von Neumann 
direct integral 
\begin{displaymath}
\mathcal{H} =\int_{\mathcal{M}} \mathcal{H}_{p} \, d\mu(p)
\end{displaymath}
of Hilbert spaces $\mathcal{H}_{p}$, compare the next section for definition.
But this is possible if the relevant quantum algebra say $\mathcal{A}$ acting
in the direct integral $\mathcal{H}$ is decomposable over a diagonal
(commutative) algebra say $\mathcal{A}^{CCR}_{1}$, such that the spectrum of
the diagonal algebra is the classical spacetime $\mathcal{M}$. Thus we arrive
at the following interpretation of the canonical field quantization:

\vspace{1ex}
{\itshape Already in the classic
works of Dirac ({\sl loc. cit.}) 
and Heisenberg ({\sl The Physical Principles of 
the Quantum Theory, Dover, New York, 1949}), 
spacetime coordinates are interpreted as representing 
a special kind of operators, called \emph{c-numbers}. The spacetime
coordinate operators $x^{\mu}\boldsymbol{1}$ commute
with each other and all other operators of our quantum algebra 
$\mathcal{A}$, which we do not specify here. Here we propose to treat 
this interpretation seriously. Let us consider an algebra 
$\mathfrak{a}$ of functions $f(p)$ on spacetime $\mathcal{M}$ 
which encodes the geometry of $\mathcal{M}$. At the moment, 
the exact structure of $\mathfrak{a}$ is not important for us. 
However, it will be a commutative algebra with point-wise operations. 
For simplicity, we confine ourself to the topological structure
of $\mathcal{M}$ (i.e. the algebra $\mathfrak{a}$ of continuous 
functions), assume $\mathcal{M}$ to be compact and 
$\mathcal{A} \subset \mathfrak{B}(\mathcal{H})$.    
Let us form an operator  $D_{f}  = f(p)\boldsymbol{1}$ 
corresponding to the function $p\to f(p)$ in our algebra 
$\mathfrak{a}$. It is therefore natural to assume that the set 
of all $D_{f}, f\in \mathfrak{a}$ in a natural 
manner composes  an algebra $\mathfrak{A}$ isomorphic to $\mathfrak{a}$, 
i.e. the point-wise function multiplication in $\mathfrak{a}$ corresponds 
to the operator composition in $\mathfrak{A}$. This is precisely what we 
mean when assuming the spacetime coordinates to be classical.
Moreover, it is also natural to assume that $\mathfrak{A}$ is closed
with respect to norm and dense in center of $\mathcal{A}$ with respect
to the strong operator topology.   
But this is possible if the operators in $\mathcal{A}$ are decomposable
and act in a direct integral of Hilbert spaces $\mathcal{H}_{p}$, 
$p\in \mathcal{M}$, or in the appropriate set of cross sections of a
Hilbert bundle $\mathcal{M}\triangle \mathcal{H}$ over spacetime}. 

But the interpretation of the inner products of the Hilbert spaces
$\mathcal{H}_{p}$, $p\in \mathcal{M}$ and of their direct integral 
$\mathcal{H}$ is not clear now. One has to preserve an open mind
on this subject. Now we know only that these Hilbert spaces 
are of importance but their role depends on the respective 
regime (remember, please, our alternative). The problem should be 
investigated in a subsequent research.  

The Dirac's critique ({\itshape loc. cit.}) inspired Piron
in his research who formulated some related ideas cf. ({\itshape C. Piron, 
physics/0204083}). He brings about within his analysis of Dirac's work
at the conclusion that some spectral families of Hilbert spaces
instead of of a mere Hilbert space are indispensable. Piron's reasoning 
was, however, completely different.

\vspace{1cm}

\begin{footnotesize}  
{\bf Appendix.} It should be mentioned an independent argument 
which shows that the generalized ray representations may play a role in QED. 
Paradoxically, there should be no zero mass vector particles with 
helicity = 1, as a consequence of the theory of unitary 
representations of the Poincar\'e group, as shown by {\L}opusza\'nski 
({\itshape Fortschritte der Physik {\bf 26}, 
261, (1978);  Rachunek spinor\'ow, PWN, Warszawa 1985 (in Polish)}). 
This is apparently in contradiction to the experiment, 
because the photon is a vector particle with helicity = 1. 
What is the solution of this paradox?
First, let us describe the solution on the grounds of the 
existing theory, which constitutes at the same time an orthodox view.
We observe that we can build a zero mass vector state with $h=1$
but we must admit finite-dimensional irreducible and thus 
\emph{non-unitary} representations of the small group, that
is the two-dimensional non-compact Euclidean group. Next, please note
that the representation of $G$ induced by the non-unitary representation
of the small group remains "unitary" if we admit the inner
product in the "Hilbert" space to be \emph{not positively} defined,
cf. ({\itshape S. N. Gupta, Proc. Phys. Soc. {\bf 63}, 681, (1950); 
K. Bleuler, Helv. Phys. Acta {\bf 23}, 567, (1950)}), or  
({\itshape S. Weinberg, The Quantum Theory of Fields, volume II, Univ. 
Press, Cambridge 1996}). 
However, even with the most favorable attitude toward the orthodox view,
this solution is rather obscure. We propose to proceed in
another way. First, let us observe that 
the case of the free quantum vector field\footnote{This
time $A_{\mu}(x)$ is an operator-valued distribution} 
$A_{\mu}(x)$ with zero mass 
is exactly the same. As long as the inner product in the Hilbert space
is positively defined, we are not able to introduce any 
vector potential which transforms as a vector field. However,
we can introduce a local real electromagnetic field
$F_{\mu\nu}(x)= -F_{\nu\mu}(x)$ which is a linear combination of a 
self-dual and an antiself-dual field with helicity $+1$ and $-1$ 
respectively. If we introduce a vector potential $A_{\mu}$
in some Lorentz frame such that 
\begin{displaymath}
F_{\mu\nu}=\partial_{\mu}A_{\nu} - \partial_{\nu}A_{\mu}
\end{displaymath}
then in another Lorentz frame we will have ($\Lambda_{\mu}^{\nu}$ is the Lorentz
transformation matrix corresponding to the Poincar\'e transformation $r$)
\begin{equation}\label{A}
A_{\mu}(x) \to U_{r}A_{\mu}(x){U_{r}}^{-1} 
= {(\Lambda^{-1})}_{\mu}^{\nu}A_{\nu}(r^{-1}x) 
+ \partial_{\mu}R(r, x), \,\,\, \partial_{\mu}R \neq 0,
\end{equation}
while $F_{\mu\nu}$ transforms as a tensor field.
We infer that gauge transformation of the second kind
has to accompany the Poincar\'e transformation, or that
gauge freedom is indispensable in the construction of the 
vector potential in the quantum field theory, cf. 
({\itshape J. {\L}opusza\'nski, Fortschritte der Physik {\bf 26}, 
261, (1978); {\sl Rachunek spinor\'ow}, PWN, Warszawa 1985 (in Polish).}) or 
({\itshape S. Weinberg, {\sl The Quantum Theory of Fields}, volume II, Univ. 
Press, Cambridge 1996}) , vol. I. 
Let $\Omega$ denote the vacuum state. According to QFT,
we should define a photon state
$\psi_{\mu}(x)$ in the following way
\begin{displaymath}
\psi_{\mu}(x)=A_{\mu}(x)\Omega.
\end{displaymath}
It immediately follows from Eq. (\ref{A}) that
\begin{equation}\label{Uphot}
U_{r}\psi_{\mu}(x)= {(\Lambda^{-1})}_{\mu}^{\nu}\psi_{\nu}(r^{-1}x) 
+ \partial_{\mu}\Theta(x),
\end{equation}
where $\partial_{\mu}\Theta(x)$ denotes the vector-valued 
distribution $\partial_{\mu}R(x)\Omega$. The above 
representation spanned by the generalized vectors $\psi_{\mu}(x)$
induces a representation in the appropriate Hilbert space.
Indeed, let us write $\varphi_{\mu}(x)$ and $\theta(x)$ for 
test functions which "smear" the distributions $\psi_{\mu}(x)$ 
and $\Theta(x)$ respectively. From formula (\ref{Uphot})
we get the transformation law for $\varphi_{\mu}$
\begin{equation}\label{Tphot}
T_{r}\varphi_{\mu}(x) = 
{(\Lambda^{-1})}_{\mu}^{\nu}\varphi_{\nu}(r^{-1}x)
+ \partial_{\mu}\theta(x).
\end{equation}
By construction, the space of test functions is dense in
the corresponding Hilbert space and the above representation
$T_{r}$ can be uniquely extended. As we are dealing with a 
gauge-invariant theory, the two quantum vector potentials $A_{\mu}(x)$ 
and $A_{\mu} + \partial_{\mu}\Phi(x)$ are unitary
equivalent. Accordingly, two photon states differing by a gradient,
as well as their two corresponding vectors $\varphi_{\mu}(x)$ and 
$\varphi_{\mu}(x) + \partial_{\mu}\phi(x)$ should be unitary 
equivalent. This means that it is more adequate to consider all 
$\varphi_{\mu}(x) + \partial_{\mu}\phi(x)$ instead of the respective
$\varphi_{\mu}(x)$ alone. We write $\varphi_{\mu}(x)+\partial_{\mu}\phi(x)$
as a pair $\{\phi(x), \varphi_{\mu}(x)\}$. The action of our 
representation $T_{r}$ in the space of pairs $\{\phi(x),\varphi_{\mu}(x)\}$ 
is as follows
\begin{displaymath}
T_{r}\{\phi(x),\varphi_{\mu}(x)\} = \{\phi(r^{-1}x)+\theta(r,x), 
U_{r}\varphi_{\mu}(x)\},
\end{displaymath}
where $U_{r}$ acts as an ordinary vector transformation:
\begin{displaymath}
U_{r}\varphi_{\mu}(x) = {(\Lambda^{-1})}_{\mu}^{\nu}\varphi_{\nu}(r^{-1}x). 
\end{displaymath}
Moreover, we have
\begin{equation}\label{QED}
T_{r}T_{s}\{\phi(x),\varphi_{\mu}(x)\} = T_{rs}\{\phi(x),\varphi_{\mu}(x)\}
+\{\xi(r,s,x),0\},
\end{equation}
where 
\begin{displaymath}
\xi(r,s,x) = \theta(r,x) + \theta(s,r^{-1}x) - \theta(rs,x).
\end{displaymath}
But this seems to be a kind of a (generalized) ray representation of the
Poincar\'e group $G$ fulfilling (\ref{rowX}) with the 
spacetime-dependent exponent $\xi(r,s,x)$! 
Summing up the discussion, we have just
seen that the generalized representation 
in the sense of Eq. (\ref{rowX}) seems to be indispensable if 
we wish to work with ordinary Hilbert spaces with positive 
norms while having a theory which describes photons.
\end{footnotesize}

\section{Generalization of Bargmann's Theory}

We have shown that we are forced to generalize the Bargmann's theory of factors to 
embrace the spacetime-dependent factors of representations acting 
in a Hilbert bundle over space-time (time) and apply the 
theory in the simplest non-relativistic case. We have already 
done it, the results will be presented in the subsequent part 
of this Chapter.

\subsection{Generalized Wave Rays and Operator Rays}\label{generalization}

In this section we give strict mathematical definitions of the
notions of the preceding section, and formulate the problem
stated there in an exact way. From the pure mathematical point 
of view, the analysis of spacetime-dependent $\xi(r,s,p)$ is more general, 
so at the outset we confine ourselves to this case \footnote{It becomes 
clear in further analysis that the group $G$ in question has 
to fulfill the consistency condition requiring that for any 
$r\in G$, $rt$ is a function of time only in the case of  
non-relativistic theory with (\ref{rowt})}. 

\vspace{1ex}

Let us recall some definitions, cf. e.g. ({\itshape G. W. Mackey, 
Unitary Group Representations in Physics, Probability, and Number Theory. 
Addison-Wesley Publishing Company, INC. New York, Amsterdam,
Wokingham-UK (1989)}). Let $\mathcal{M}$ be a set endowed with an analytic 
Borel structure. 

\vspace{1ex}
 
By a \emph{Hilbert bundle over} $\mathcal{M}$ or a \emph{Hilbert bundle
with base} $\mathcal{M}$ we shall mean an assignment 
$\mathcal{H}: p \to\mathcal{H}_{p}$ of a Hilbert space $\mathcal{H}_{p}$
to each $p \in \mathcal{M}$. The set of all pairs $(p,\psi)$ with
$\psi \in \mathcal{H}_{p}$ will be denoted by 
$\mathcal{M}\triangle\mathcal{H}$ and called the \emph{space of the bundle}.  
By a \emph{cross section} of our bundle we shall mean an assignment
$\psi: p \to \psi_{p}$ of a member of $\mathcal{H}_{p}$ to each 
$p\in\mathcal{M}$. If $\psi$ is a cross section and $(p_{0},\phi_{0})$
a point of $\mathcal{M}\triangle\mathcal{H}$, we may form a scalar product
$(\phi_{0},\psi_{p_{0}})$. In this way, every cross-section $\psi$ defines a 
complex-valued function $f_{\psi}$ on $\mathcal{M}\triangle\mathcal{H}$. 
By a \emph{Borel Hilbert bundle} we shall mean
a Hilbert bundle  together with an analytic Borel structure in
$\mathcal{M}\triangle\mathcal{H}$ such that the following conditions 
are fulfilled
\begin{enumerate}
\item[(1)]  Let $\pi(p,\psi)=p$. Then $E \subseteq \mathcal{M}$ is a Borel 
set if and only if $\pi^{-1}(E)$ is a Borel set in 
$\mathcal{M}\triangle\mathcal{H}$.

\item[(2)]  There exist countably many cross-sections 
$\psi^{1}, \psi^{2}, \ldots$ such that 
\begin{itemize}
\item[(a)] the corresponding complex-valued functions on 
$\mathcal{M}\triangle\mathcal{H}$ are Borel functions,

\item[(b)] these Borel functions separate points in the sense 
that no two distinct points $(p_{i},\phi_{i})$ of 
$\mathcal{M}\triangle\mathcal{H}$ assign the same values to all
$\psi^{j}$ unless $\phi_{1}=\phi_{2}=0$, and

\item[(c)] $p \to (\psi^{i}(p),\psi^{j}(p))$ is a Borel function for all 
$i$ and $j$.
\end{itemize}
\end{enumerate}

A cross-section is said to be a \emph{Borel cross-section} 
if the function on $\mathcal{M}\triangle\mathcal{H}$ defined by the 
cross-section is a Borel function. All Borel cross-sections
compose a linear space under the obvious operations, cf. Mackey 
({\itshape loc. cit}). Now let $\mu$ be a measure on $\mathcal{M}$. 
The cross-section $p \to \varphi_{p}$ is said to be \emph{square summable} 
with respect to $\mu$ if 
\begin{displaymath}
\int_{\mathcal{M}} (\varphi_{p}, \varphi_{p}) \, d\mu(p) < \infty.
\end{displaymath} 
The space $\mathcal{L}^{2}(\mathcal{M}, \mu, \mathcal{H})$  of 
all equivalence classes of square-summable cross-sections, where two
cross-sections $\varphi$ and $\varphi'$ are in the same equivalence 
class if $\varphi_{p}=\varphi'_{p}$ for almost all $p \in \mathcal{M}$,
forms a separable Hilbert space with the inner product given by
\begin{displaymath}
(\varphi, \theta) = \int_{\mathcal{M}} (\varphi_{p}, \theta_{p}) \, d\mu(p),
\end{displaymath}
cf. Mackey ({\itshape loc. cit.}). It is called the direct integral of the $\mathcal{H}_{p}$
with respect to $\mu$ and is denoted by 
$\int_{\mathcal{M}} \mathcal{H}_{p} \, d\mu(p)$. 

\vspace{1ex}

Identification with the previous section is partially suggested by the 
notation itself. We shall make this identification more explicit. 
The set $\mathcal{M}$ plays the role of spacetime or real 
line $\mathbb{R}$  of time $t$ respectively. The wave functions  
$\psi$ of the preceding section are the Borel cross-sections of 
$\mathcal{M}\triangle\mathcal{H}$ but if they do belong to the  
subset $\mathcal{L}^{2}(\mathcal{M},\mu,\mathcal{H})$ of cross-sections 
which are square integrable presents an open question. The separate Hilbert spaces 
$\mathcal{H}_{p}$ with their inner products play some role in experiments 
as well as the inner product in their direct integral product. 
But would be better to leave unspecified the precise role they play 
in experiments for now.
We have also 
used  $\psi(p)$ and $\psi_{p}$ as well as 
$(\psi_{p}, \theta_{p})$ and $(\psi, \theta)_{p}$ interchangeably.

\vspace{1ex}

By an \emph{isomorphism} of the Hilbert bundle 
$\mathcal{M}\triangle\mathcal{H}$ with the Hilbert bundle
$\mathcal{M}'\triangle\mathcal{H}'$ we shall mean a Borel isomorphism $T$
of $\mathcal{M}\triangle\mathcal{H}$ on $\mathcal{M}'\triangle\mathcal{H}'$ 
such that for each $p \in \mathcal{M}$ the restriction of $T$ to
$p \times \mathcal{H}_{p}$ has some $q \times \mathcal{H}_{q}'$
for its range and is unitary when regarded as a map of $\mathcal{H}_{p}$
on $\mathcal{H}_{q}'$. The induced map carrying $p$ into $q$ is clearly
a Borel isomorphism of $\mathcal{M}$ with $\mathcal{M}'$ and we denote it
by $T^{\pi}$. The above-defined $T$ is said to be an \emph{automorphism}
if $\mathcal{M}\triangle\mathcal{H}=\mathcal{M}'\triangle\mathcal{H}'$. 
Please note that for any automorphism $T$ we have 
$(T\psi,T\phi)_{T^{\pi}p}=(\psi,\phi)_{p}$, but in general
$(T\psi,T\phi)_{p}\neq(\psi,\phi)_{p}$. By this token, any automorphism 
$T$ is what is frequently called \emph{bundle isometry}. 

The function $r \to T_{r}$ from group $G$ into the set of automorphisms
(bundle isometry) of $\mathcal{M}\triangle\mathcal{H}$ is said to be
a \emph{general factor representation} of $G$ associated 
to the action $G \times \mathcal{M} \ni r,p \to r^{-1}p \in \mathcal{M}$ of
$G$ on $\mathcal{M}$ if $T_{r}^{\pi}(p) \equiv r^{-1}p$ for all $r \in G$,
and $T_{r}$ satisfy condition (\ref{rowX}). 

\vspace{1ex}

Of course, $T_{r}$ is to be identified with that of the preceding
section. Our further specializing assumptions partly following from the above
interpretation are as follows. We assume $\mathcal{M}$ to be endowed with 
the manifold structure inducing a topology associated with the 
above-assumed Borel structure. We confine ourselves to a finite dimensional
Lie group $G$ which acts smoothly and transitively on spacetime 
$\mathcal{M}$, such that a $G$-invariant measure $\mu$ exists on $\mathcal{M}$. 

\vspace{1ex}

By a \emph{factor representation} of a Lie group we mean a general
factor representation with the exponent $\xi(r,s,p)$ differentiable in 
$p \in \mathcal{M}$.

Now we define the \emph{operator ray} $\boldsymbol{T}$ corresponding
to a given bundle isometry operator $T$ to be the set of operators 
\begin{displaymath}
{\boldsymbol{T}}=\{\tau T ,  p \to \tau(p) \in {\mathcal{D}} \, \, 
\textrm{and} \, \,  \vert\tau \vert = 1\},
\end{displaymath} 
where $\mathcal{D}$ denotes the set of all differentiable 
real functions on $\mathcal{M}$.  
Any $T\in {\boldsymbol{T}}$ will be called a \emph{representative} of the 
ray ${\boldsymbol{T}}$. The product ${\boldsymbol{TV}}$ is defined as the 
set of all products $TV$ such that $T\in {\boldsymbol{T}}$ and 
$V \in {\boldsymbol{V}}$. 

\vspace{1ex}

Please note that not all Borel sections are physically realizable.
By interpreting the discussion of the preceding section in
the Hilbert bundle language, we see that the role of the 
Schr\"odinger equation is essentially to establish all
the physical sections. Any two sections $\psi(p)$ and
$\psi'(p) =e^{i\zeta(p)}\psi(p)$ are indistinguishable
giving the same probabilities $\vert f_{\psi}\vert^{2}
=\vert f_{\psi'}\vert^{2}$. After this, any group
$G$ acting in $\mathcal{M}$ induces a \emph{ray representation} of $G$,
i.e. a mapping $r \to \boldsymbol{T}_{r}$ of $G$ into the space of rays of
bundle automorphisms (bundle isometrics) of 
$\mathcal{M}\triangle\mathcal{H}$, fulfilling the condition
\begin{displaymath}
\boldsymbol{T}_{r}\boldsymbol{T}_{s}=\boldsymbol{T}_{rs}.
\end{displaymath}
For any cross-section $\psi$ we define its corresponding 
\emph{ray} $\boldsymbol{\psi} =\{e^{i\zeta(p)}\psi(p), 
\zeta \in \mathcal{D}\}$. If $\psi$ is a physical cross-section, then we get 
the \emph{physical ray} of the preceding section. 
Selecting a representative $T_{r}$ for each $\boldsymbol{T}_{r}$, we get
a factor representation fulfilling (\ref{rowX}).
Please note that $T_{r}$ transforms rays 
into rays, and we have $T_{r}(e^{i\xi(p)}\psi)=e^{i\xi_{r}(p)}T_{r}\psi$. 
Further on we assume that that  operators
$T_{r}$ are such that   $\xi_{r}(p)=\xi(r^{-1}p)$, where $r^{-1}p$ denotes 
the action of $r^{-1} \in G$ on the spacetime
point $p \in \mathcal{M}$. This is a natural assumption 
which does actually take place in practice.    

Now we shall make the last assumption, namely that all transition 
probabilities vary continuously with
the continuous variation of the coordinate transformation $s \in G$:
\begin{enumerate}
\item[] For any element $r$ in $G$, any ray ${\boldsymbol{\psi}}$ 
and any positive $\epsilon$, there exists a neighborhood $\mathfrak{N}$ of $r$ 
on $G$ such that 
$d_{p}(\boldsymbol{T}_{s}\boldsymbol{\psi},
\boldsymbol{T}_{r}\boldsymbol{\psi}) < \epsilon$ if 
$s \in \mathfrak{N}$ and $p \in \mathcal{M}$,
\end{enumerate}
where
\begin{displaymath}
d_{p}(\boldsymbol{\psi_{1}},\boldsymbol{\psi_{2}}) = 
\inf_{\psi_{i} \in \boldsymbol{\psi}_{i}}
\Vert \psi_{1} - \psi_{2} \Vert_{p} 
= \sqrt{2\vert1-\vert(\psi_{1},\psi_{2})_{p}\vert \, \vert}.
\end{displaymath}

Basing on the continuity assumption, one can prove the following

\vspace{1ex}

\begin{twr} Let ${\boldsymbol{T}}_{r}$ be a continuous ray representation of 
a group $G$. For all $r$ in a suitably chosen neighborhood $\mathfrak{N}_{0}$ 
of the unit element $e$ of $G$ one may select a strongly continuous set of 
representatives $T_{r}\in {\boldsymbol{T}}_{r}$. That is, for any compact set 
$\mathcal{C} \subset \mathcal{M}$, any wave function $\psi$, 
any $r \in {\mathfrak{N}}_{0}$ and any positive $\epsilon$ there exists a 
neighborhood $\mathfrak{N}$ of $r$ such that $\Vert T_{s}\psi 
- T_{r}\psi \Vert_{p} < \epsilon$ if $s \in \mathfrak{N}$ and 
$p \in \mathcal{C}$. \end{twr}     

\vspace{1ex}

There are numerous possible selections of such factor representations.
But many among them merely differ by a differentiable
phase factor and are physically indistinguishable. 
We call them equivalent. Our task then is to classify all 
possible factor representations with respect 
to this equivalence.

\subsection{Local Exponents}\label{exponents}

The representatives $T_{r} \in \boldsymbol{T}_{r}$ selected as in Theorem
 2 will be called \emph{admissible}, with the representation $T_{r}$ obtained 
in this way referred to as an \emph{admissible} representation. 
There are infinitely many 
possibilities of such a selection of admissible representations
$T_{r}$. We confine ourselves to the local \emph{admissible} representations 
defined on a fixed neighborhood $\mathfrak{N}_{o}$ of $e \in G$, as in 
Theorem 2. 

Let $T_{r}$ be an \emph{admissible} representation. With the help of the 
phase $e^{i\zeta(r,p)}$ with a real function $\zeta(r,p)$ differentiable in 
$p$ and continuous in $r$, we can define
\begin{equation}\label{4}
T'_{r} = e^{i\zeta(r,p)}T_{r},
\end{equation}
which is a new \emph{admissible} representation. This is trivial if one 
defines the continuity of $\zeta(r,p)$ in $r$ appropriately. Namely, 
from Theorem 2 it follows that the continuity has to be defined
in the following way. The function $\zeta(r,p)$ \emph{will be called strongly 
continuous in r at $r_{0}$ if and only if for any compact set $\mathcal{C} 
\subset \mathcal{M}$
 and any positive $\epsilon$ there exists a 
neighborhood ${\mathfrak{N}}_{0}$ of $r_{0}$ such that } 
\begin{displaymath} 
\vert \zeta(r_{0},p) - \zeta(r,p) \vert < \epsilon,
\end{displaymath}
\emph{for all} $r \in {\mathfrak{N}}_{0}$ \emph{and for all} 
$p \in \mathcal{C}$. But the converse is also true. Indeed,
if $T'_{r}$ is also an \emph{admissible} representation, then (\ref{4}) 
has to be fulfilled for a real function $\zeta(r,p)$
differentiable in $p$ because $T'_{r}$ and $T_{r}$ belong to the same ray. 
Moreover, because both $T'_{r}\psi$ and $T_{r}\psi$ are strongly 
continuous (in $r$ for any $\psi$), then $\zeta(r,p)$ has to be 
\emph{strongly continuous} (in $r$).

Let $T_{r}$ be an \emph{admissible} representation, and thus continuous 
in the sense indicated in Theorem 2. One can always choose the above 
$\zeta$ in such a way that $T_{e} = 1$ as will be assumed from now on.

Because $T_{r}T_{s}$ and $T_{rs}$ belong to the same ray, one has
\begin{equation}\label{5}
T_{r}T_{s} = e^{i\xi(r,s,p)} T_{rs}
\end{equation}    
with a real  function $\xi(r,s,p)$  differentiable in $p$. From the fact that 
$T_{e} = 1$, we have
\begin{equation}\label{9}
\xi(e,e,p) = 0.
\end{equation}  
From the associative law $(T_{r}T_{s})T_{g} = T_{r}(T_{s}T_{g})$ one gets
\begin{equation}\label{10}
\xi(r,s,p) + \xi(rs,g,p) = \xi(s,g,r^{-1}p) + \xi(r,sg,p).
\end{equation} 
Formula (\ref{10}) is very important and our analysis largely rests on 
this relation. From the fact that the representation $T_{r}$ is 
\emph{admissible} follows that 
the exponent $\xi(r,s,p)$ is continuous in 
$r$ and $s$. Indeed, let us take a $\psi$ belonging
to a unit ray $\boldsymbol{\psi}$. Then, making use of (\ref{5}), we get
\begin{displaymath}
e^{i\xi(r,s,p)}(T_{rs} - T_{r's'})\psi + (T_{r'}(T_{s'} - T_{s})\psi + 
(T_{r'} - T_{r})T_{s}\psi 
\end{displaymath}  

\begin{displaymath}
= (e^{i\xi(r',s',p)} - e^{i\xi(r,s,p)}) T_{r's'}\psi.
\end{displaymath}
Taking norms $\Vert \centerdot \Vert_{p}$ of both sides, we get 
\begin{displaymath}
\vert e^{i\xi(r',s',p)} - e^{i\xi(r,s,p)} \vert \leq 
\Vert (T_{r's'} - T_{rs})\psi \Vert_{p} + 
\end{displaymath}

\begin{displaymath}
+ \Vert T_{r'}(T_{s'} - T_{s})\psi \Vert_{p} + \Vert (T_{r'} - 
T_{r})T_{s}\psi \Vert_{p}.
\end{displaymath} 
From this inequality and the continuity of $T_{r}\psi$, the continuity of 
$\xi(r,s,p)$ in $r$ and $s$ follows.
Moreover, from Theorem 2 and the above inequality follows the 
\emph{strong continuity} of 
$\xi(r,s,p)$ in $r$ and $s$.  

Formula (\ref{4}) suggests the following definition. Two 
\emph{admissible} representations
$T_{r}$ and $T'_{r}$ are called \emph{equivalent} if and only if 
$T'_{r} = e^{i\zeta(r,p)}T_{r}$ 
for some real function $\zeta(r,p)$ 
differentiable in $p$ and \emph{strongly continuous} in $r$. Thus, making use 
of 
(\ref{5}), we get $T'_{r}T'_{s} = e^{i\xi'(r,s,p)}T'_{rs}$, where
\begin{equation}\label{13}
\xi'(r,s,p) = \xi(r,s,p) + \zeta(r,p) + \zeta(s,r^{-1}p) - \zeta(rs,p).
\end{equation}
Then the two exponents $\xi$ and $\xi'$ are equivalent if and only if 
(\ref{13}) is fulfilled with
 $\zeta(r,p)$ \emph{strongly continuous} in 
$r$ and differentiable in $p$. 

From (\ref{9}) and (\ref{10}) it immediately follows that
\begin{equation}\label{11}
\xi(r,e,p) = 0 \, \, \, and \, \, \, \xi(e,g,p) = 0,
\end{equation}

\begin{equation}\label{12}
\xi(r,r^{-1},p) = \xi(r^{-1},r,r^{-1}p).
\end{equation}
Relation (\ref{13}) between $\xi$ and $\xi'$ will be written in short by 
\begin{equation}\label{13'}
\xi' = \xi + \Delta[\zeta].
\end{equation}
The relation (\ref{13}) between exponents $\xi$ and $\xi'$ defines an 
equivalence relation, which preserves the linear structure.

We introduce now group $H$, a very important notion for our further 
investigations. It is evident
 that all operators $T_{r}$ contained in all 
rays $\boldsymbol{T}_{r}$ form a group under multiplication.
Indeed, let us consider an \emph{admissible} representation $T_{r}$ with a 
well-defined $\xi(r,s,p)$ in formula (\ref{5}).
Because any $T_{r} \in \boldsymbol{T}_{r}$ has the form $e^{i\theta(p)}T_{r}$ 
(with a real and differentiable $\theta$), one has
\begin{equation}\label{H-action}
\Big( e^{i\theta(p)}T_{r}\Big) \Big( e^{i\theta'(p)}T_{s}\Big) = 
e^{i\{\theta(p) + \theta'(r^{-1}p) + \xi(r,s,p)\}}T_{rs}.
\end{equation}
This important relation suggests the following definition of the local group 
$H$ connected with the \emph{admissible} representation or with the exponent 
$\xi(r,s,p)$. Namely, $H$ consists of the pairs $\{\theta(p), r\}$ where 
$\theta(p)$ is a differentiable real function and $r \in G$. The 
multiplication rule, suggested by the above relation, is defined as follows
\begin{equation}\label{15}
\{\theta(p),r\} \centerdot \{\theta'(p),r'\} = \{\theta(p) + 
\theta'(r^{-1}p) + \xi(r,r',p), \, rr' \}.
\end{equation}
The associative law for this multiplication rule is equivalent to (\ref{10}) 
(in complete analogy with the classical Bargmann's theory). The pair 
$\check{e} =\{0,e\}$ plays the role of the unit element in $H$. For any 
element $\{\theta(p), r\} \in H$ there exists the inverse 
$\{\theta(p),r\}^{-1} = \{-\theta(rp) - \xi(r,r^{-1},rp), \, r^{-1} \}$. 
Indeed, from (\ref{12}) it follows that $\{\theta, r\}^{-1} \centerdot 
\{\theta,r \} = \{\theta, r\} \centerdot \{\theta, r\}^{-1} = \check{e}$. 
The elements
 $\{\theta(p), e\}$ form an Abelian subgroup $N$ of $H$. Any 
$\{\theta,r\} \in H$ can be uniquely written as
 $\{\theta(p),r\} = 
\{\theta(p),e\} \centerdot \{ 0, r\}$. The same element can be also uniquely 
expressed in the form $\{\theta(p),r\} = \{0,r\} \centerdot \{\theta(rp),e\}$. 
Thus, we have $H = N \centerdot G = G \centerdot N$. The Abelian subgroup $N$ 
is a normal factor subgroup of $H$. But this time, $G$ does not form any 
normal factor subgroup of $H$ (contrary to the classical case investigated 
by Bargmann, when the exponents do not depend on $p$). So, this time $H$ is 
not direct product $N \otimes G$, but a semidirect 
product $N \circledS G$. In this case, however, the theorem that $G$ 
is locally isomorphic to the factor group $H/N$ is still valid. 
Then group $H$ composes a \emph{semicentral extension} 
of $G$, and not a central extension of $G$ as in the Bargmann's theory.

The rest of this paper is based on the following reasoning (the author 
being largely inspired by Bargmann's work ({\itshape Ann. Math. {\bf 59}, 
1, 1954}). If the two exponents 
$\xi$ and $\xi'$ are \emph{equivalent}, that is $\xi' = \xi + \Delta[\zeta]$, 
then the \emph{semicentral extensions} $H$ and $H'$ connected
with $\xi$ and $\xi'$ are homomorphic. The homomorphism $h: \{\theta,r\} 
\mapsto \{\theta', r'\}$ is given by 
\begin{equation}\label{izo}
\theta'(p) = \theta(p) - \zeta(r,p), \, \, r' = r.
\end{equation}
 Using an \emph{Iwasawa-type construction} we show that any exponent 
$\xi(r,s,p)$ is equivalent to a differentiable one (in $r$ and $s$). 
We can then confine ourselves to the differentiable $\xi$ and $\xi'$. 
We show  that  $\zeta(r,p)$ is also a differentiable function of $(r,p)$. 
Moreover, we show that any $\xi$ is equivalent to the canonical one, 
that is such $\xi$ which is differentiable and for which 
$\xi(r,s,p) = 0$ whenever $r$ and $s$ 
belong to the same one-parameter subgroup.
Then we can restrict our investigation to the canonical $\xi$  
considering the subgroup of all elements $\{\theta(p),r\} \in H$ with 
differentiable $\theta(p)$. For simplicity let us denote the subgroup 
by the same symbol $H$ . We embed the subgroup in an infinite dimensional Lie 
group $D$ with manifold structure modeled on a Banach space. Then we consider the 
subgroup $\overline{H}$ which is the closure of $H$ in $D$. After this,
$\overline{H}$ turns into a Lie group and the homomorphism (\ref{izo})
becomes an isomorphism of the two Lie groups. Thus, the group 
$\overline{H}$ has the Banach Lie algebra $\overline{\mathfrak{H}}$.   
We apply the general theory of analytic groups developed in 
({\itshape G. Birkhoff, Continuous Groups and Linear Spaces, Recueil 
Math\'ematique (Moscow) {\bf 1}(5), 635, (1935); Analytical Groups, Trans. 
Am. Math. Soc. {\bf 43}, 61, (1938)}) and ({\itshape E. Dynkin, Uspekhi 
Mat. Nauk {\bf 5}, (1950), 135; Amer. Math. Soc. Transl. {\bf 9}(1), 
(1950), 470}). From  this theory it follows that the correspondence between the 
local $\overline{H}$ and $\overline{\mathfrak{H}}$ 
is bi-unique and one can construct uniquely the local group $\overline{H}$
from the algebra $\overline{\mathfrak{H}}$ as well. As we will see, 
the algebra defines a spacetime-dependent anti-linear form 
$\Xi$ on the Lie algebra $\mathfrak{G}$ of $G$, the so-called 
\emph{infinitesimal exponent} $\Xi$. By this we reduce the classification
of local $\xi$'s which define $\overline{H}$'s to the classification 
of $\Xi$'s which define $\overline{\mathfrak{H}}$'s. So, we will simplify 
the 
problem of the classification of local $\xi$'s to a largely linear 
problem. Here are the details.

\vspace{1ex} 

{\bf Iwasawa construction}. Let us denote by ${\ud}r$ and ${\ud}^{*}r$ the 
left and right invariant Haar measure on $G$. Let $\nu(r)$ and $\nu^{*}(r)$ be 
two infinitely differentiable functions on $G$ with compact supports contained 
in the fixed neighborhood $\mathfrak{N}_{0}$ of $e$. By multiplying them by the 
appropriate constants, we can always obtain: $\int_{G} \nu(r) \, {\ud}r = 
\int_{G} \nu^{*}(r) \, {\ud}^{*}r =1$. Let $\xi(r,s,p)$ be any
\emph{admissible} local exponent defined on $\mathfrak{N}_{0}$. We will 
construct a differentiable (in $r$ and $s$) exponent $\xi''(r,s,p)$  
\emph{equivalent} to $\xi(r,s,p)$ and defined on $\mathfrak{N}_{0}$, in 
the following two steps: $\xi' = \xi + \Delta[\zeta]$ and
$\xi'' = \xi' + \Delta[\zeta']$, where  
$\zeta(r,p)$ is the left invariant integral of $l\to -\xi(r,l,p)\nu(l)$, while
$\zeta'(r,p)$ is the right invariant integral of 
$u\to -\xi'(u,r,up)\nu^{*}(u)$.
A rather simple computation in which we use (\ref{13}) and (\ref{10}) 
and the invariance property of the Haar measures 
shows
that $\xi''(r,s,p)$ is a differentiable 
(up to any order)exponent in all variables.
Next we shall show that \emph{if two differentiable exponents 
$\xi$ and $\xi'$ are equivalent, that is, if $\xi' = \xi + \Delta[\zeta]$, 
then $\zeta(r,p)$ is differentiable in $r$}. Clearly, the difference $\xi'-\xi$
is differentiable. Similarly, both $(\xi'-\xi)\nu$  ( with $\nu$
defined as above), as well as its left invariant integral $\eta$ are 
differentiable. It is easy to show that $\zeta' = \eta - \zeta$ is also 
differentiable. In this way we arrive at differentiability of
$\zeta = \eta - \zeta'$ is differentiable. A slightly more
complicated argumentation shows that \emph{every (local) exponent of 
one-parameter group is equivalent to zero}. However, the argumentation is quite 
analogous to that of Bargmann. We can treat such a group as the additive 
group of real numbers, so that the first two arguments of $\xi$ are real 
numbers. Let us set $\vartheta(\tau,\sigma,p)$ as the derivative of $\xi$ 
with respect to the second argument. It is not hard to show that 
$\xi + \Delta[\zeta]=0$, where
$\zeta(\tau,p)$ is the ordinary Riemann integral of 
$\mu\to \tau\vartheta(\mu\tau,0,p)$ over the unit interval $[0,1]$. But it
means that $\xi$ is equivalent to zero. 

Let us recall that the continuous curve $r(\tau)$ in a Lie group $G$ is 
a one-parameter subgroup if and only if $r(\tau_{1})r(\tau_{2}) = 
r(\tau_{1} + \tau_{2})$, \emph{i.e.} $r(\tau) = (r_{0})^{\tau}$ for
some element $r_{0} \in G$. (Please note that the real power $r^{\tau}$ 
is well defined on a Lie group, at least
on some neighborhood of $e$). The coordinates $\rho^{k}$ in $G$ 
are \emph{canonical} if and only if any curve of the form 
$r(\tau) = \tau \rho^{k}$ (where the coordinates $\rho^{k}$ are 
fixed) is a one-parameter subgroup 
(the curve $r(\tau) = \tau \rho^{k}$ 
will be denoted in short by $\tau a$, with the coordinates 
of $a$ equal to $\rho^{k}$). The "vector" $a$ is called by 
physicists the \emph{generator} of the one-parameter subgroup $\tau a$. 

A local exponent $\xi$ of a Lie group $G$ is called \emph{canonical} if 
$\xi(r,s,p)$ is differentiable in all variables, and $\xi(r,s,p) = 0$ 
if $r$ and $s$ are elements of the same one-parameter subgroup.

Almost the same argument used to show that every 
$\xi$ on a one-parameter group is equivalent to zero also shows that 
\emph{every local exponent $\xi$ of a Lie group is equivalent 
to a canonical local exponent}. In order to prove this, we shall apply
the argument to the exponent $\xi_{0}(\tau, \sigma,p)
:=\xi(\tau a, \sigma a,p)$, cf. ({\itshape J. Wawrzycki, math-ph/0301005}).  
Up to now, the argumentation has been more or less analogous to that
of Bargmann. From now on, the argumentation becomes entirely different. 
\emph{Let $\xi$ and $\xi'$ be two differentiable and equivalent 
local exponents of a Lie group $G$, assuming $\xi$ to 
be canonical. Then $\xi'$ is canonical if and only if $\xi' = \xi 
+ \Delta[\Lambda]$, where $\Lambda(r,p)$ is a linear form in the 
canonical coordinates of $r$ fulfilling the condition that
$\Lambda(a,(\tau a)p)$ is constant as a function of $\tau$,
i.e. it follows that}\footnote{The limit in the expression 
can be understood in the ordinary 
point-wise sense with respect to $p$, but also in any linear topology in the
function linear space (with obvious addition) of $\theta(p)$,  
providing that $p \to \Lambda(a, p)$ is differentiable in the sense 
of this linear topology. Further on, the simple notation 
\begin{displaymath}
\boldsymbol{a} f(p) = \frac{df((\tau a)p)}{d\tau}\Big\vert_{\tau = 0}
= \lim_{\epsilon \to 0}\frac{f((\epsilon a)p) - f(p)}{\epsilon},
\end{displaymath}
will be used.}
\begin{equation}\label{condLem}
\boldsymbol{a}\Lambda(a,p) = \frac{d\Lambda(a, (\tau a)p)}{d\tau} 
= \lim_{\epsilon \to 0}\frac{\Lambda(a,(\epsilon a)p) 
- \Lambda(a,p)}{\epsilon} = 0.
\end{equation}
While sufficiency of the condition in the above statement is almost 
evident, proving its necessity is quite nontrivial. 
Hereafter we outline the argumentation. Because 
the exponents are equivalent we have 
$\xi'(r,s,p) = \xi(r,s,p) +\Delta[\zeta]$. Since both $\xi$ and 
$\xi'$ are differentiable then $\zeta(r,p)$ is also a differentiable 
function, which follows from what has been said above.  Let us 
suppose that $r=\tau a$ and $s= \tau' a$. Because both $\xi$ and $\xi'$ 
are \emph{canonical}, we have 
$\xi(\tau a, \tau' a,p) = \xi'(\tau a, \tau' a,p) = 0$, so
that $\Delta[\zeta](\tau a, \tau' a,p) = 0$. Applying the last formula 
recurrently one gets
\begin{displaymath}
\zeta(\tau a,p) = \sum_{k=0}^{n-1} \zeta(\frac{\tau}{n}a,
(-\frac{k}{n}\tau a)p).
\end{displaymath}
Then we use the Taylor Theorem to each summand in the above expression, 
and pass to the limit $n \to + \infty$. In this way, we obtain 
\begin{equation}\label{18}
\zeta(\tau a,p) = \int_{0}^{\tau} \varsigma(a,(-\sigma a)p) \, {\ud} \sigma,
\end{equation}
where $\varsigma = \varsigma(r,p)$ is a differentiable function, 
cf. ({\itshape J. Wawrzycki, math-ph/0301005}). If we differentiate now 
expression (\ref{18}) with respect to $\tau$ at $\tau = 0$, 
we will immediately see that the function 
$\varsigma(a,p)$ is linear with respect to $a$. Let us suppose that the 
spacetime coordinates are chosen in such a way that the integral 
curves $p(x) = (xa)p_{0}$ are coordinate lines, 
which is possible for appropriately small $x$. There are of course 
three remaining families of coordinate lines besides $p(x)$, which can be 
chosen in an arbitrary way, with their parameters denoted by $y_{i}$.
After this, 
\begin{displaymath} 
\zeta(a,x,y_{i}) = \frac{1}{\tau} \int_{0}^{\tau} \varsigma(a,x 
- \sigma, y_{i}) \, {\ud} \sigma 
= \frac{1}{\tau} \int_{x - \tau}^{x} \varsigma(a,z,y_{i}) \, {\ud}z,
\end{displaymath}
for any $\tau$ (of course, with appropriately small $\vert \tau \vert$, 
in our case $\vert \tau \vert \leq 1$) and for any (appropriately small) 
$x$. But this is possible for the function $\varsigma(a,x,y_{k})$ 
continuous in $x$  (in our case, differentiable in $x$) 
if and only if $\varsigma(a,x,y_{k})$ does not depend on $x$. This means 
that $\zeta(a,x,y_{k})$ does not depend on $x$ and the condition of the 
statement is hereby proved.

\vspace{1ex}

{\bf Infinitesimal exponents and embedding of $H$ in a Lie group $D$}.
According to what has been shown already, we can assume that the 
exponent is canonical. We also confine ourselves to the subgroup of 
$\{\theta(p),r\} \in  H$ with differentiable $\theta$, and denote 
this subgroup by the same letter $H$. We embed this subgroup $H$ in an 
infinite dimensional Lie group with the manifold structure modeled on 
a Banach space. We will extensively use the theory developed by Birkhoff 
({\itshape loc. cit.}) and Dynkin ({\itshape loc. cit.}). For the systematic 
treatment of manifolds modeled on Banach spaces, see e.g. ({\itshape S. Lang, 
Differential Manifolds, Springer-Verlag, Berlin, Heidelberg, New York (1985)}).  
By this embedding we ascribe bi-uniquely a Lie algebra to the group $H$ with
the convergent Baker-Hausdorff series.  

Please note first that the formula
\begin{displaymath}
H \times \mathcal{L}^{2}(\mathcal{M},\mu,\mathcal{H})\ni 
(\{\theta(p),r \}, \phi) \to e^{i\theta(p)}T_{r}\phi
\end{displaymath}
(together with (\ref{H-action})) can be viewed as a rule
giving the action of $H$ in the direct integral Hilbert space
$\int_{\mathcal{M}} \mathcal{H}_{p} \, d\mu(p)$ defined in  
Section {\bf 3}. Moreover, this is a unitary action, provided
$\mu$ is $G$-invariant. In accordance to Birkhoff ({\itshape loc. cit.}), 
the group $D$ of \emph{all} unitary operators of a Hilbert space is an infinite 
dimensional Lie group. Hence, $H = N \circledS G$ can be viewed as a 
subgroup of a Lie group. 

We consider now the closure $\overline{H}$ of $H$ in the sense of the 
topology in $D$. It is remarkable that \emph{the subgroup 
$\overline{H}$ has locally the structure of the semi-direct 
product $\overline{N} \circledS G$ as well}. This is a consequence of the 
following four facts. (1) $\overline{N}$ is a normal subgroup of 
$\overline{H}=\overline{N\centerdot G}$. (2) $G$ is finite dimensional, 
so $\overline{G} = G$. (3) Locally (in a neighborhood $\mathcal{O}$), 
the multiplication in $D$ is given by the Baker-Hausdorff formula in the 
Banach algebra of $D$. Because $\overline{N}$ is  normal in $\overline{H}$, 
then the above exponential mapping converts locally  the multiplication 
$\overline{N}\centerdot S$ of $\overline{N}$ by any subset $S$ of 
$\overline{H}$ into the sum $\overline{N} + S$. Because $G$ is 
finite-dimensional, and hence locally compact, the  neighborhood 
$\mathcal{O}$ can be chosen in such a way that locally (in the closure 
of $\mathcal{O} + \mathcal{O}$) the following holds:
\begin{displaymath}
\overline{N}+\overline{G} 
= \overline{N+G} = \overline{H}.
\end{displaymath}	  
(4) The local $\overline{N}$ (intersected with $\overline{\mathcal{O}}$) 
has a finite co-dimension in local $\overline{N+G}$ (intersected with
$\overline{\mathcal{O} + \mathcal{O}}$) and thus it
splits locally $\overline{N+G}$. So, we have locally \emph{i.e.} 
in $\overline{\mathcal{O} + \mathcal{O}}$:
\begin{displaymath}
\overline{N} + \overline{G}
= \overline{H} = \overline{N}\oplus G', 
\end{displaymath}
where $\overline{G'} = G'$ and  $\oplus$ stands for a direct sum.
From this it follows that $G' =\overline{G}$ locally. This shows
that $\overline{H}=\overline{N}\circledS G$.
 	  
Because $\overline{H}= \overline{N}\circledS G$, every $h\in \overline{H}$
is uniquely representable in the form $ng$, where $n \in \overline{N}$ and 
$g \in G$. Please note now that
\begin{displaymath}
(n_{1}g_{1})(n_{2}g_{2}) = n_{1}g_{1}n_{2}g_{1}^{-1}g_{1}g_{2} =
[n_{1}(g_{1}n_{2}g_{1}^{-1})](g_{1}g_{2})
\end{displaymath}
and that $g_{1}n_{2}g_{1}^{-1} \in \overline{N}$ because $\overline{N}$
is normal in $\overline{H}$. Let us denote the automorphism 
$n \to gng^{-1}$ of $\overline{N}$ by $R_{g}$. 
The group $\overline{H}$ can be locally viewed as a topological product of 
Banach spaces $\overline{\mathfrak{N}}\times 
\mathfrak{G}$,  one of which (namely $\mathfrak{G}$) is
finite-dimensional and isomorphic tho the Lie algebra of $G$.
The multiplication in $\overline{H}$ can be written as
$(n_{1}, g_{1})(n_{2},g_{2}) = (n_{1}R_{g_{1}}(n_{2}), g_{1}g_{2})$. 
Moreover, $\overline{N}$ can be viewed locally as the Banach space 
$\overline{\mathfrak{N}}$ with the multiplication law given by the vector 
addition in $\overline{\mathfrak{N}}$.   

Our task now is to reconstruct the Lie algebra $\overline{\mathfrak{H}}$
corresponding to the subgroup $\overline{H}$. Let $\lambda \to \lambda a$ 
be a one-parameter subgroup of $G$.
The mapping $(\lambda, n) \to (R_{\lambda a}n, \lambda a)$
of the Banach space $\mathcal{R} \times \overline{\mathfrak{N}}$ 
into the Banach space $\overline{\mathfrak{N}} \times \mathfrak{G}$ is 
continuous. In consequence, $\mathcal{R}\ni \lambda \to R_{\lambda a}n \in 
\overline{\mathfrak{N}}$ as well as $\overline{\mathfrak{N}} 
\ni n \to R_{\lambda a}n$ are continuous. Therefore, the function 
$\lambda \to R_{\lambda a}n$
can be integrated over any compact interval and  
\begin{displaymath}
\tau \to (n_{\tau a}, \tau a) 
:= \Big(\int_{0}^{\tau}R_{\sigma a}n \, \ud \sigma, \tau a\Big) 
\end{displaymath}
is a one-parameter subgroup of $\overline{H}$ with 
generator\footnote{The limit process with the help of which 
the generator is computed refers to the topology in $D$, 
of course.} $(n, a)$, cf. Birkhoff ({\itshape loc. cit.}), Dynkin 
({\itshape loc. cit}). 
Having obtained this, we are able to reconstruct the algebra. 
The elements of $H \subset \overline{H}$ are representable in the 
ordinary form $\{\alpha,r\}$ with 
differentiable $\alpha = \alpha(p), p\in \mathcal{M}$, and $r \in G$. 
Let us consider the above-defined operator $R_{\lambda a}$. Its restriction
to $H \subset \overline{H}$ is given by (please remember that $\xi$ is canonical) 
\begin{displaymath}
\alpha(p) \to (R_{\lambda a}\alpha)(p) = \alpha((\lambda a)^{-1}p).
\end{displaymath}
We can now compute explicitly the Lie bracket and the Jacobi identity
for all the elements $\{\alpha(p), a\}$ of the subalgebra $\mathfrak{H} 
\subset \overline{\mathfrak{H}}$ corresponding to the subgroup $H$.    
The result is as follows\footnote{Let us stress once more that  
\begin{equation}\label{strong}
\boldsymbol{a}\theta(p) = \lim_{\epsilon \to 0}\frac{\theta((\epsilon a)p) 
- \theta(p)}{\epsilon},
\end{equation}
and the limit is in the sense of topology induced from the Lie group $D$.}
\begin{equation}\label{20'}
[\check{a}, \check{b}] = \{\boldsymbol{a}\beta - \boldsymbol{b}\alpha 
+ \Xi(a,b, p), [a,b] \},
\end{equation}
  
\begin{displaymath}
\Xi(a,b,p) = \lim_{\tau \to 0} \tau^{-2}\{\xi((\tau a)(\tau b), 
(\tau a)^{-1}(\tau b)^{-1},p) + 
\end{displaymath}

\begin{equation}\label{20''}
+ \xi(\tau a, \tau b, p) + \xi((\tau a)^{-1},(\tau b)^{-1}, 
(\tau b)^{-1}(\tau a)^{-1}p) \},
\end{equation}

From the associative law in $H$ one gets
\begin{displaymath}
\Xi([a,a'],a'',p) + \Xi([a',a''],a,p) + \Xi([a'',a],a',p) =
\end{displaymath} 

\vspace{-0.5cm}

\begin{equation}\label{21}
= \boldsymbol{a}\Xi(a',a'',p) + \boldsymbol{a'}\Xi(a'',a,p) 
+ \boldsymbol{a''}\Xi(a,a',p),
\end{equation}
which can be shown to be equivalent to the Jacobi identity
\begin{equation}\label{21'}
[[\check{a}, \check{a}'], \check{a}''] + [[\check{a}',\check{a}''],\check{a}] 
+ [[\check{a}'',\check{a}], \check{a}'] = 0. 
\end{equation}
Thus, in this way we have reconstructed the Lie algebra 
$\overline{\mathfrak{H}}$ giving explicitly $[\check{a}, \check{b}]$ for all 
$\check{a},\check{b} \in \mathfrak{H} \subseteq \overline{\mathfrak{H}}$. 
Because $\mathfrak{H}$ is dense in $\overline{\mathfrak{H}}$, the local 
exponent $\Xi$ determines the algebra $\overline{\mathfrak{H}}$ uniquely. 
But from the theory of Lie groups the correspondence between the algebras 
$\overline{\mathfrak{H}}$ and local Lie groups $\overline{H}$ is bi-unique, 
at least locally, cf. e.g. Birkhoff ({\itshape loc. cit.}) and Dynkin
({\itshape loc. cit.}). Therefore,  
\emph{the correspondence} $\overline{H} \to 
\overline{\mathfrak{H}}$ \emph{between the local group} $\overline{H}$ 
\emph{and the algebra} $\overline{\mathfrak{H}}$ \emph{is one-to-one}.
Because the exponent $\xi$ determines the multiplication rule in $H$ and 
vice-versa, then it follows that \emph{the correspondence $\xi \to \Xi$ 
between the local $\xi$ and the infinitesimal exponent $\Xi$ is one-to-one}.
Please note that the term 'local $\xi = \xi(r,s,p)$' means that $\xi(r,s,p)$ is 
defined for $r$ and $s$ belonging to a fixed neighborhood 
${\mathfrak{N}}_{0} \subset G$ of $e \in G$, but 
\emph{in our case it is defined globally as a function of the 
spacetime variable $p \in \mathcal{M}$}.

\vspace{1ex}

{\bf Infinitesimal exponents and local exponents}.
Now, let us move on to describing the relation between the infinitesimal 
exponents $\Xi$ and the local exponents $\xi$. First, let us compute the 
infinitesimal exponents $\Xi$ and $\Xi'$ given by (\ref{20''}),
which correspond to the two equivalent canonical local exponents $\xi$ and 
$\xi' = \xi + \Delta[\Lambda]$. Inserting $\xi' = \xi + \Delta[\Lambda]$ 
into formula (\ref{20''}), one gets
\begin{equation}\label{24}
\Xi'(a,b,p) = \Xi(a,b,p) +  \boldsymbol{a}\Lambda(b,p) 
- \boldsymbol{b}\Lambda(a,p) - \Lambda([a,b],p). 
\end{equation}     
According to what has been said, we can confine ourselves to the 
canonical exponents. Then, as one of our previous statements
said, $\Lambda = \Lambda(a, (\tau b)p)$ 
is a constant function of $\tau$ if $a = b$, and $\Lambda(a,p)$ is linear 
with respect to $a$ (we use the canonical coordinates on $G$). 
Hence $\Xi'(a,b,p)$ is antisymmetric in $a$ and $b$ and fulfills 
(\ref{21}) only if $\Xi(a,b,p)$ is antisymmetric in $a$ and $b$ and fulfills 
(\ref{21}). This suggests the definition: \emph{two infinitesimal exponents 
$\Xi$ and $\Xi'$ will be called equivalent if and only if relation} 
(\ref{24}) \emph{holds}. For brevity, we write relation (\ref{24}) 
as follows: 
\begin{displaymath}
\Xi' = \Xi + d[\Lambda].
\end{displaymath}
Finally, we maintain that \emph{two canonical local exponents $\xi$ and 
$\xi'$ are equivalent if and only if the corresponding infinitesimal 
exponents $\Xi$ and $\Xi'$ are equivalent}. Indeed. 
(1) Assume $\xi$ and $\xi'$ to be equivalent. Then, by the 
definition of equivalence of infinitesimal
exponents: $\Xi' = \Xi + d[\Lambda]$. (2) Assume $\Xi$ and $\Xi'$ to be 
equivalent: $\Xi' = \Xi + d[\Lambda]$ for some linear form $\Lambda(a,t)$ 
such that $\Lambda(a,(\tau a)p)$ does not depend on $\tau$. 
Then $\xi + \Delta[\Lambda] \to \Xi'$, and by the uniqueness of the 
correspondence $\xi \to \Xi$ we have $\xi' = \xi + \Delta[\Lambda]$, 
\emph{i.e.} $\xi$ and $\xi'$ are equivalent. In this way, we arrive at
the following:

\vspace{1ex}

\begin{twr} (1) On a Lie group G, every local exponent $\xi(r,s,p)$ 
is equivalent to a canonical local exponent $\xi'(r,s,p)$  which, 
on some canonical neighborhood ${\mathfrak{N}}_{0}$, is analytic
in canonical coordinates of r and s, and vanishes if r and s belong 
to the same one-parameter subgroup. Two canonical
local exponents $\xi,\xi'$ are equivalent if and only if 
$\xi' = \xi + \Delta[\Lambda]$ on some canonical
neighborhood, where $\Lambda(r,p)$ is a linear form in the canonical 
coordinates of $r$ such that $\Lambda(r,sp)$ does not depend on $s$
if $s$ belongs to the same one-parameter subgroup as $r$.
(2) To every canonical
 local exponent of $G$ there corresponds uniquely an 
infinitesimal exponent $\Xi(a,b,p)$ on the Lie
 algebra $\mathfrak{G}$ of $G$, 
i.e. a bilinear antisymmetric form which satisfies  the identity 
$\Xi([a,a'],a'',p) +\Xi([a',a''],a,p)+ \Xi(a'',a],a',p) 
= \boldsymbol{a}\Xi(a',a'',p) + \boldsymbol{a'}\Xi(a'',a,p) +
\boldsymbol{a''}\Xi(a,a',p)$. The correspondence is linear. (3) Two canonical 
local exponents 
$\xi,\xi'$ are equivalent if and only if the corresponding 
$\Xi,\Xi'$ are equivalent, i.e. $\Xi'(a,b,p) = \Xi(a,b,p) 
+ \boldsymbol{a}\Lambda(b,p) - \boldsymbol{b}\Lambda(a,p) - \Lambda([a,b],p)$
where $\Lambda(a,p)$ is a linear form in $a$ on $\mathfrak{G}$ such that 
$\tau \to \Lambda(a,(\tau b)p)$ is constant if $a = b$.
(4) There exist a one-to-one correspondence between the equivalence classes 
of local exponents $\xi$ (global in $p \in \mathcal{M}$) of $G$ and the 
equivalence classes of infinitesimal exponents $\Xi$ of 
$\mathfrak{G}$. \end{twr}

\subsection{Global Extensions of Local Exponents}

Theorem 3 provides full classification of exponents $\xi(r,s,p)$ local 
in $r$ and $s$, defined for all $p \in \mathcal{M}$.
But if $G$ is both connected and simply connected,  
then we have the following theorems. (1) If an extension $\xi'$ of a given 
local (in $r$ and $s$) exponent $\xi$ does exist, then it is uniquely 
determined (up to the equivalence transformation (\ref{13})) (Theorem 4). 
(2) There exists such an extension $\xi'$ (Theorem 5),
proved for $G$, which possess finite-dimensional extension
$\mathfrak{H}'$ only.

We are not able to prove that the (\emph{global}) homomorphism
(\ref{izo}) is continuous when $\xi$ is not canonical. Please note 
that any $\xi$ is equivalent to its canonical counterpart, but only 
\emph{locally}! This is why the topology of $H$ induced from $D$ 
is not applicable in the global analysis. 
We introduce another topology. Because of the semidirect structure
We introduce another topology. Because of the semidirect structure
of $H = N\circledS G$, it is sufficient to introduce it into
$N$ and $G$ separately in such a manner that $G$ acts continuously 
on $N$, cf. e.g. Mackey ({\itshape loc. cit.}). From the discussion of Section
\ref{exponents} it is sufficient to introduce the Fr\'echet
topology of almost uniform convergence in the function space $N$. 
Indeed, from the strong continuity of $\xi$ and $\zeta$ in
(\ref{izo}) it follows that the multiplication rule as well as
the homomorphism (\ref{izo}) are continuous. 

\vspace{1ex}

\begin{twr} Let $\xi$ and $\xi'$ be two equivalent local exponents of a 
connected and simply connected group $G$, so that $\xi' = \xi 
+ \Delta[\zeta]$ on some neighborhood, assuming the exponents 
$\xi_{1}$ and $\xi_{1}'$ of $G$ to be extensions of $\xi$ and $\xi'$ 
respectively. Then, for all $r,s \in G$:
$\xi_{1}'(r,s,p) = \xi_{1}(r,s,p) + \Delta[\zeta_{1}]$, where $\zeta_{1}(r,p)$ 
is strongly
 continuous in $r$ and differentiable in $p$, and $\zeta_{1}(r,p) 
= \zeta(r,p)$, for all $p \in \mathcal{M}$ and for all 
$r$ belonging to some neighborhood of $e \in G$. \end{twr}
 
\vspace{1ex}  

Here is the proof outline. The two exponents $\xi_{1}$ and $\xi_{1}'$ being 
\emph{strongly continuous} (by assumption) 
define two semicentral extensions $H_{1} = N_{1} \circledS G$ and 
$H_{1}'= N'_{1}\circledS G$, which are continuous groups. 
Please note that the linear groups $N_{1},N'_{1}$ are connected and 
simply connected. Because both $H_{1}$ and $H_{1}'$
are semi-direct products of two connected
and simply connected groups they are both connected and simply connected. 
Eq. (\ref{izo}) defines a local isomorphism mapping $h: \check{r} \to 
\check{r}' = h(\check{r})$ of $H_{1}$ into $H_{1}'$.
Because $H_{1}$ and $H_{1}'$ are 
connected and simply connected, the isomorphism $h$ given by (\ref{izo}) can 
be uniquely extended to an isomorphism 
$h_{1}$ of the entire groups 
$H_{1}$ and $H_{1}'$ such 
that $h_{1}(\check{r}) = h(\check{r})$ on some neighborhood of $H_{1}$, cf.
({\itshape L. Pontrjagin, {\sl Topological groups}, Moscow (1984) 
(in Russian)}), Theorem 80. The isomorphism $h_{1}$ defines
an isomorphism of the two Abelian subgroups  $N_{1}$ and $h_{1}(N_{1})$. 
By (\ref{izo}), $h_{1}(\theta,e) = \{\theta, e\}$ locally in $H_{1}$, that is 
for $\theta$ lying appropriately 
close to 0 (in the metric sense defined 
previously). Both $N_{1}$ and $h_{1}(N_{1})$ are connected, and $N_{1}$ is 
in addition simply connected,
so applying once again Theorem 80 of Pontrjagin ({\itshape loc. cit.}), one can see that
$h_{1}(\theta,e) = \{\theta,e\}$ for all $\theta$. A rather simple computation
shows that $\zeta_{1}$ defined by the equality $h_{1}(0,r) 
= \{ - \zeta_{1}(p),g(r)\}$ fulfills the conditions of our theorem.

\vspace{1ex}

The following theorem is proved for the group $G$ with a finite-dimensional
extended algebra $\mathfrak{H}'$.

\vspace{1ex}

\begin{twr} Let $G$ be a connected and simply connected Lie group. Then to
every exponent $\xi(r,s,X)$ of $G$ defined locally in $(r,s)$ there exists an exponent 
$\xi_{0}$ of $G$ defined on the whole group $G$ which is an extension of $\xi$. If $\xi$
is differentiable, $\xi_{0}$ may be chosen differentiable. \end{twr}

\vspace{1ex}

Because the proof of Theorem 5 is almost identical to that of Theorem 5.1 in 
Bargmann ({\itshape loc. cit.}), we do not present it explicitly\footnote{In this 
proof we consider the finite-dimensional extension $H'$ of $G$ instead of the 
Lie group $H$ in the proof presented in Bargmann ({\itshape loc. cit.}). 
The remaining replacements are rather trivial, but we 
mark them here explicitly to simplify the reading 
(1) Instead of the formula $\bar{r}' = \bar{t}(\theta)\bar{r} 
= \bar{r}\bar{t}(\theta)$ of  (5.3) in Bargmann ({\itshape loc. cit.}), we have 
$\check{r}' = \check{t}(\theta(r^{-1}p))\check{r} 
= \check{r}\check{t}(\theta(p))$. 
Thus, from the formula $(\check{h}_{1}(r)\check{h}_{1}(s))\check{h}_{1}(g)$ 
$= \check{h}_{1}(r)(\check{h}_{1}(s)\check{h}_{1}(g))$ see
{\itshape V. Bargmann, loc. cit.}) it 
follows that $\xi(r,s,p) + \xi(rs,g,p)$ $= \xi(s,g,r^{-1}p) + \xi(r,sg,p)$ 
instead of (5.8) in Bargmann ({\itshape loc. cit.}). (2) Instead of (4.9), 
(4.10) and (4.11) we use the Iwasawa-type construction presented in this paper.}. 
Please note that the proof rests largely on the global theory of classical 
(finite-dimensional) Lie groups. Namely, it rests on the theorem that
to any finite dimensional Lie group there always exists a universal 
covering group . We can use those methods because of the existence
of a finite-dimensional extension $H'$ of $G$.

We have obtained the full classification of time-dependent $\xi$ 
defined on the whole group $G$ for Lie groups $G$ which are connected 
and simply connected in non-relativistic
theory.  But for any Lie group $G$ there exists
a universal covering group $G^{*}$ which is connected and simply connected. 
Thus, for $G^{*}$ the correspondence $\xi \to \Xi$ is one-to-one, that is, 
to every $\xi$ there exists a unique $\Xi$
 and vice versa, to every $\Xi$ 
corresponds a unique $\xi$ defined on the whole group $G^{*}$, and the 
correspondence preserves the equivalence relation. Because $G$ and $G^{*}$ 
are locally isomorphic, the infinitesimal exponents $\Xi$'s are exactly the 
same for $G$ and for $G^{*}$. Since to every $\Xi$ there does exist 
exactly one $\xi$ on $G^{*}$, so, if to a given $\Xi$ there exists the 
corresponding $\xi$ on the whole $G$, then such a $\xi$ is unique.
In this way, we have obtained the full classification of $\xi$ defined
on a whole Lie group $G$ for any Lie group $G$, in the sense that no $\xi$
 can be omitted in the classification. The set of equivalence classes of 
$\xi$ is considerably smaller than that for $\Xi$; 
it may happen that to some local $\xi$ there does not exist any global 
extension.

\subsection{Examples}\label{examples}

{\bf Example 1: The Galilean Group}.
According to the conclusions of subsection \ref{Time} one should 
\emph{a priori} investigate such representations of 
the Galilean group $G$ 
which fulfill Eq. (\ref{rowt}), with $\xi$ depending on time. 
Then, the following paradox arises. Why has the transformation law $T_{r}$ under 
the Galilean group a time-independent $\xi$ in (\ref{rowt}), regardless of 
whether it is a covariance group or a symmetry group? We will solve the 
paradox in this subsection. Namely, we will show that any representation
of the Galilean group fulfilling (\ref{rowt}) is equivalent to a 
representation fulfilling (\ref{rowt}) with time-independent $\xi$. 
This is a rather peculiar property of the Galilean group, not valid in 
general. For example, this is not true for the group of Milne transformations.

In  non-relativistic theory $\xi=\xi(r,s,t)$ depends on the time. 
In this case, according to our assumption about $G$, any $r \in G$ 
transforms simultaneity hyperplanes into simultaneity hyperplanes. Thus, 
there are two
possibilities for any $r \in G$. First, when $r$ does not 
change time: $t(rp) = t(p)$, and the second in which time is 
changed, but in such a way that $t(rp) - t(p) = f(t)$. \emph{We assume 
in addition that the base generators $a_{k} \in \mathfrak{G}$ can be 
chosen in such a way that only one acts on time as translation
and the remaining ones do not act on time}. 
We can assume that the operators $\boldsymbol{a}$ are ordinary 
differential operators. Hence, the Jacobi identity (\ref{21}) reads
\begin{equation}\label{Jac1}
\Xi([a,a'],a'') + \Xi([a', a''],a) + \Xi([a'',a],a') 
= \partial_{t} \Xi(a',a''),
\end{equation}
if one and only one among $a,a',a''$ is the time-translation generator, 
namely $a$, and 
\begin{eqnarray}\label{Jac2}
\Xi([a,a'],a'') + \Xi([a',a''], a) + \Xi([a'',a],a') = 0,
\end{eqnarray} 
in all of the remaining cases. The Jacobi identities (\ref{Jac1}) and (\ref{Jac2}) 
can be treated as a system of ordinary differential linear equations for 
the finite set of unknown functions $\Xi_{ij}(t) = \Xi(a_{i}, a_{j},t)$, 
where $a_{i}$ is the base in the Lie algebra of $G$.

According to Section \ref{generalization}, in order to classify all $\xi$ of $G$
we shall determine all equivalence classes of infinitesimal exponents 
$\Xi$ of the Lie algebra $\mathfrak{G}$ of 
$G$. The commutation relations for the Galilean 
group are as follows
\begin{equation}\label{26a}
[a_{ij},a_{kl}] = \delta_{jk}a_{il} - \delta_{ik}a_{jl}+\delta_{il}a_{jk}-
\delta_{jl}a_{ik}, 
\end{equation}

\begin{equation}\label{26b}
[a_{ij},b_{k}] = \delta_{jk}b_{i} - \delta_{ik}b_{j}, \, [b_{i},b_{j}] = 0,
\end{equation}

\begin{equation}\label{26c}
[a_{ij.d_{k}}] = \delta_{jk}d_{i} - \delta_{ik}d_{j}, \, [d_{i},d_{j}] 
= 0, [b_{i},d_{j}] = 0,
\end{equation}

\begin{equation}\label{26d}
[a_{ij},\tau] = 0, [b_{k},\tau] = 0, [d_{k},\tau] = b_{k},
\end{equation}
where $b_{i},d_{i}$ and $\tau$ stand for the generators of space translations, 
with the proper Galilean transformations and time translation respectively and 
$a_{ij} = - a_{ji}$ being rotation generators. Please note that the Jacobi identity 
(\ref{Jac2}) is identical to that in the ordinary
Bargmann's theory of time-independent exponents see Bargmann ({\itshape loc. cit.}), 
Eqs (4.24) and (4.24a)). Thus, using (\ref{26a}) -- (\ref{26c}) we can proceed exactly 
after Bargmann ({\itshape loc. cit.}), pages 39, 40) and show that any infinitesimal 
exponent 
defined on the subgroup generated by $b_{i}, d_{i}, a_{ij}$ is 
equivalent to an exponent
equal to zero, with the possible exception of 
$\Xi(b_{i},d_{k},t) = \gamma \delta_{ik}$, where $\gamma = \gamma(t)$. Hence, 
the only components of $\Xi$ defined on the whole algebra $\mathfrak{G}$ which 
can be a priori not equal to zero are: 
$\Xi(b_{i},d_{k},t) = \gamma \delta_{ik}, \, \Xi(a_{ij}, 
\tau,t), \,
\Xi(b_{i}, \tau,t)$ and $\Xi(d_{k},\tau,t)$. First, we compute 
function $\gamma(t)$. Substituting $a= \tau, \, a' = b_{i}, a'' = d_{k}$ 
to (\ref{Jac1}), we get ${\ud}\gamma/{\ud}t = 0$, so that $\gamma$
is a constant, denoting the constant value of $\gamma$ by $m$. By inserting 
$a = \tau, \, a' = a_{i}^{s}, \, a'' = a_{sj}$ to (\ref{Jac1}) and summing 
up with respect to $s$, we get $\Xi(a_{ij},\tau,t) = 0$. 
In the same way, but with the substitution $a = \tau, a' = a_{i}^{s}, a'' 
= b_{s}$, one shows that
$\Xi(b_{i},\tau, t) = 0$. At last, the substitution 
$a=\tau, a' = a_{i}^{s}, a''=d_{s}$ to (\ref{Jac1})
and summation with respect to $s$ gives $\Xi(d_{i},\tau,t) = 0$. 
In this way, we have proved that any time-dependent $\Xi$ on $\mathfrak{G}$ 
is equivalent to a time-independent one. In other words, we get a 
one-parameter family of possible $\Xi$, with the parameter equal to the 
inertial mass $m$ of the system in question. Any infinitesimal time-dependent 
exponent  of the Galilean group is equivalent to the above time-independent 
exponent $\Xi$ with some value of the parameter $m$; and any two infinitesimal
exponents with different values of $m$ are nonequivalent.   
As was argued in subsection \ref{generalization} (Theorems 3 $\div$ 5), the 
classification of $\Xi$ gives a full classification of $\xi$.
Moreover, it can be shown that the classification of $\xi$ 
is equivalent to the classification of possible $\theta$-s 
in the transformation law
\begin{equation}\label{trapsi}
T_{r}\psi(p) = e^{i\theta(r,p)}\psi(r^{-1}p)
\end{equation}
for the spinless non-relativistic particle.
On the other hand, the exponent $\xi(r,s,t)$ of the representation $T_{r}$ 
given by (\ref{trapsi}) can be easily computed to be equal to
$\theta(rs,p) - \theta(r,p) - \theta(s,r^{-1}p)$,
and the infinitesimal exponent belonging to $\theta$ defined as 
$\theta(r,p) = -m\vec{v}\centerdot \vec{x} + \frac{m}{2}\vec{v}^{2}t$,
covers the whole one-parameter family of the classification 
(its infinitesimal exponent is equal to that infinitesimal exponent $\Xi$,
which has been found above). Thus, the standard $\theta(r,p) = - m\vec{v} 
\centerdot \vec{x} + \frac{m}{2}\vec{v}^{2}t$, covers the full 
classification of possible $\theta$-s in (\ref{trapsi}) for the Galilean 
group. Inserting the standard form for $\theta$ we see that $\xi$ does not 
depend on time but only on $r$ and $s$. By this, any time-dependent
$\xi$ on $G$ is equivalent to its time-independent counterpart. 

In this way, we have reconstructed the standard result. Using now the 
formula\footnote{The transformation 
$T_{r}$ does not act in the ordinary Hilbert 
space but in the Hilbert bundle space $\mathcal{R}\triangle\mathcal{H}$,
hence we cannot immediately appeal to the Stone and 
G{\aa}rding Theorems. Nonetheless, $T_{r}$ induces a unique unitary 
representation acting in the Hilbert space 
$\int_{\mathcal{R}} \mathcal{H}_{t} d\mu(t)$ 
and it can be shown that it is meaningful to talk 
about the generators $A$ of $T_{r}$.} 
\begin{displaymath}
A_{i}\psi(p) = \lim_{\sigma \to 0} \frac{(T_{\sigma a_{i})} 
- \boldsymbol{1})\psi(p)}{\sigma},
\end{displaymath}
for the generator $A_{i}$ corresponding to $a_{i}$,
we get the standard commutation relations for the \emph{ray} 
representation $T_{r}$ of the Galilean group 
\begin{displaymath}
[A_{ij},A_{kl}] = \delta_{jk}A_{il} - \delta_{ik}A_{jl} - \delta_{jl}A_{ik},  
\end{displaymath}

\begin{displaymath}
[A_{ij},B_{k}] = \delta_{jk}B_{i} - \delta_{ik}B_{j}, \, [B_{i},B_{j}] = 0,
\end{displaymath}

\begin{displaymath}
[A_{ij}, D_{k}] = \delta_{jk}D_{i} - \delta_{ik}D_{j}, 
\end{displaymath}

\begin{displaymath}
[D_{i},D_{j}] =0, \, [B_{i},D_{j}] = m\delta_{ij},
\end{displaymath}

\begin{displaymath}
[A_{ij}, T] = 0, \, [B_{k},T] = 0, \, [D_{k}, T] = B_{k}.
\end{displaymath} 

Please note that to any $\xi$ (or $\Xi$) there exists a corresponding $\theta$ 
(and  such a $\theta$ is unique up to a trivial equivalence relation). 
As we will see, this is not the case for the Milne group, where some 
$\Xi$'s do exist which cannot be realized by any $\theta$. 
 
\vspace{1ex}

{\bf Example 2: Milne group and equality of inertial and gravitational masses}. 
In here 
we apply the theory of Section \ref{generalization} to 
the Milne transformations group. We proceed like with the Galilean group 
in the preceding section. The Milne group $G$ does not form
any Lie group, but in the physical application it is sufficient for us 
to consider some Lie subgroups $G(m)$ of the Milne group.  
We will go on according 
to the following plan. First, we compute the infinitesimal exponents
and exponents for each $G(m)$, 
$m = 1,2, \ldots $, and then the $\theta$ in (\ref{trapsi}) for $G(m)$.
Please compare ({\itshape J. Wawrzycki, math-ph/0301005}), where the result 
is extended on the whole group.

The Milne transformation is defined as follows  
\begin{equation}\label{tra}
(\vec{x},t) \to (R\vec{x} + \vec{A}(t), t + b),
\end{equation}
where $R$ is an orthogonal matrix, and $b$ is constant. 
The extent of arbitrariness of function 
$\vec{A}(t)$ in (\ref{tra}) will be left undetermined for now. 
It is convenient to rewrite the Milne transformations (\ref{tra}) in the 
following form
\begin{displaymath}
\vec{x'} = R\vec{x} + A(t) \vec{v}, \, \, \, t' = t + b,
\end{displaymath}
where $\vec{v}$ is a constant vector which does not depend on time $t$. 
We define
the subgroup $G(m)$ of $G$ as the group of the following 
transformations
\begin{displaymath}
\vec{x'} = R\vec{x} + \vec{v}_{(0)} + t\vec{v}_{(1)} 
+ \frac{t^{2}}{2!}\vec{v}_{(2)} + \ldots + \frac{t^{m}}{m!}\vec{v}_{(m)},
\, \, \, t' = t + b, 
\end{displaymath}
where $R = (R_{a}^{b}), v_{(n)}^{k}$ are the group parameters; 
in particular, the group $G(m)$ has the dimension equal to $3m + 7$. 

Now Let us investigate the group $G(m)$, that is, classify its 
infinitesimal exponents. The commutation relations of $G(m)$ are 
as follows
\begin{equation}\label{Milne1}
[a_{ij},a_{kl}] = \delta_{jk}a_{il} - \delta_{ik}a_{jl} 
+ \delta_{il}a_{jk} - \delta_{il}a_{ik},
\end{equation}

\begin{equation}\label{Milne2}
[a_{ij},d_{k}^{(n)}] = \delta_{jk}d_{i}^{(n)} 
- \delta_{ik}d_{j}^{(n)}, \, [d_{i}^{(n)},d_{j}^{(k)}] = 0, 
\end{equation}

\begin{equation}\label{Milne3}
 [a_{ij},\tau] = 0, \,  [d_{i}^{(0)},\tau]=0, \, [d_{i}^{(n)},\tau] 
= d_{i}^{(n-1)}, 
\end{equation}
where $d_{i}^{(n)}$ is the generator of the transformation 
$r(v_{(n)}^{i})$:
\begin{displaymath}
{x'}^{i} = x^{i} + \frac{t^{n}}{n!}v_{(n)}^{i},
\end{displaymath}
which will be called the $n$-acceleration, and 0-acceleration in the
particular case of the ordinary space translation. 
All relations (\ref{Milne1}) and 
(\ref{Milne2}) are identical to (\ref{26a}) $\div$ (\ref{26c}) with
the $n$-acceleration instead of the Galilean transformation. Thus, the 
same argumentation as that used for the Galilean group gives: 
$\Xi(a_{ij}, a_{kl})= 0$, $\Xi(a_{ij},d_{k}^{(n)}) = 0$, and 
$\Xi(d_{i}^{(n)}, d_{j}^{(n)}) = 0$. Substituting $a_{i}^{h}, a_{hi}, 
\tau$ for $a,a',a''$ into Eq. (\ref{Jac1}),making use of the 
commutation relations and summing up with respect to $h$, we get 
$\Xi(a_{ij},\tau) = 0$. In an analogous way, substituting 
$a_{i}^{h}, d_{h}^{(l)}, d_{k}^{(n)}$ for $a,a',a''$ into  
Eq. (\ref{Jac2}), we get  $\Xi(d_{i}^{(l)}, d_{k}^{(n)}) 
= \frac{1}{3}\Xi(d^{(l)h},d_{h}^{(n)}) \, \delta_{ik}$.
Substituting $a_{i}^{h}, d_{h}^{(n)}, \tau$ for $a,a',a''$ into 
Eq. (\ref{Jac1}), making use of the commutation relations,
and summing up with respect to $h$, we get $\Xi(d_{i}^{(n)}, \tau) = 0$. 
Now, we substitute $d_{k}^{(n)}, d_{i}^{(0)}, \tau$ for $a,a',a''$ in 
(\ref{Jac1}), and proceed recurrently with respect to $n$,
obtaining in this way $\Xi(d_{i}^{(0)},d_{k}^{(n)}) 
= P^{(0,n)}(t)\delta_{ik}$, where $P^{(0,n)}(t)$ is a polynomial
of degree $n-1$, and the time derivation of $P^{(0,n)}(t)$ has to be 
equal to $P^{(0,n-1)}(t)$, and $P^{(0,0)}(t) = 0$. 
Substituting $d_{k}^{(n)}, d_{i}^{(l)}, \tau$ into (\ref{Jac1}), 
in the same way we get
$\Xi(d_{k}^{(l)}, d_{i}^{(n)}) = P^{(l,n)}(t)\delta_{ki}$, where 
$\frac{{\ud}}{{\ud}t}P^{(l,n)} = P^{(l-1,n)} + P^{(l,n-1)}$.
This allows us to determine all $P^{(l,n)}$ by the recurrent integration 
process. Please note that $P^{(0,0)} = 0$, and $P^{(l,n)} = - P^{(n,l)}$, so 
given the $P^{(0,n)}$ we can compute all $P^{(1,n)}$. Indeed, we have 
$P^{(1,0)} = - P^{(0,1)}, P^{(1,1)} = 0, {\ud}P^{(1,2)}/{\ud}t = P^{(0,2)} 
+ P^{(1,1)}, {\ud}P^{(1,3)}/{\ud}t = P^{(0,3)} + P^{(1,2)},
\ldots$, and after $m-1$ integrations we compute all $P^{(1,n)}$. Each 
elementary integration introduces a new independent parameter (the arbitrary 
additive integration constant). Exactly in the same way, given all 
$P^{(1,n)}$ we can compute all $P^{(2,n)}$ after $m-2$ elementary 
integration processes. In general, the $P^{(l-1,n)}$ allows us to compute 
all $P^{(l,n)}$ after $m-l$ integrations. Thus, $P^{(l,n)}(t)$ are 
$l+n - 1$-degree polynomial functions of $t$, and all are determined by 
$m(m+1)/2$ integration constants. Because 
$d[\Lambda](d_{i}^{(n)}, d_{k}^{(l)}) = 0$, the exponents $\Xi$ defined 
by different polynomials $P^{(l,n)}$ are inequivalent. Therefore, the space 
of nonequivalent classes of $\Xi$ is $m(m+1)/2$-dimensional.

However, not all $\Xi$ can be realized by the transformation $T_{r}$ of 
the form (\ref{trapsi}). It can seen that any integration constant 
$\gamma_{(l,q)}$ of the polynomial $P^{(l,q)}(t)$ has to be equal to zero 
if $l,q \neq 0$, provided the exponent $\Xi$ belongs to the
representation $T_{r}$ of the form (\ref{trapsi}).
By this, all exponents of $G(m)$ which can be realized by the 
transformations $T_{r}$ of the form (\ref{trapsi}) are determined by the 
polynomial $P^{(0,m)}$, that is, by $m$ constants. We omit the proof of this
fact, and refer the reader to ({\itshape J. Wawrzycki, math-ph/0301005}), 
Example 2. In the proof we compute
the exponent $\Xi$ directly for the transformation $T_{r}$ of the form 
(\ref{trapsi}) and compare it with the classification results above.
 
Consider the $\theta$, given by the formula
\begin{equation}\label{thetaG(m)}
\theta(r,p) = \gamma_{1}\frac{{\ud}\vec{A}}{{\ud}t} 
+ \gamma_{2}\frac{{\ud}^{2}\vec{A}}{{\ud}t^{2}} + \ldots + 
\gamma_{m}\frac{{\ud}^{m}\vec{A}}{{\ud}t^{m}} + \widetilde{\theta}(t),
\end{equation}
for $r \in G(m)$, where $\gamma_{i}$ are the integration constants which 
define the polynomial $P^{(0,m)} = \gamma_{1}\frac{t^{m-1}}{(m-1)!} 
+ \gamma_{2}\frac{t^{(m-2)}}{(m-2)!} + \ldots + \gamma_{m}$, and
$\widetilde{\theta}(t)$ is any function of time $t$, and eventually 
of the group parameters. A rather simple computation shows that this 
$\theta$ covers all possible $\Xi$ which can be realized by (\ref{trapsi}). 
That is, the infinitesimal exponents corresponding to the $\theta$ given by 
(\ref{thetaG(m)}) yield all possible $\Xi$ with 
all integration constants $\gamma_{(k,n)} = 0$, for $k,n \neq 0$. 
Thus, the most general $\theta(r,p)$ defined for $r \in G(m)$ is given by 
(\ref{thetaG(m)}). 

At this point we make use of the assumption that the wave equation is local. 
It can be shown then (we leave this without proof)
that the $\theta(r,p)$ can be a function of a finite order derivatives of 
$\vec{A}(t)$, say $k$-th at most, while the higher derivatives cannot enter into 
$\theta$. Therefore, the most general $\theta(r,p)$ defined for $r \in G(m)$ 
has the following form 
\begin{equation}\label{thetaG}    
\theta(r,X) = \gamma_{1}\frac{{\ud}\vec{A}}{{\ud}t} + \ldots 
+ \gamma_{k}\frac{{\ud}^{k}\vec{A}}{{\ud}t^{k}} 
+ \widetilde{\theta}(t).
\end{equation}

Having obtained this we can infer the most general Schr\"odinger
equation for a spinless particle in Newton-Cartan spacetime,
cf. ({\itshape J. Wawrzycki, Acta Phys. Polon. {\bf B 35}, 613 (2004); 
gr-qc/0301102.}). The inertial and the gravitational masses
are always equal in this equation.  
Our assumptions are, more precisely, as follows: (i) The quantum particle,
when its kinetic energy is small in comparison to its rest energy $mc^{2}$,
does not exert any influence on the space-time structure. (ii) The Born 
interpretation for the wave function is valid, and the transition 
probabilities in the Newton-Cartan space-time which describes geometrically
Newtonian gravity, are equal to the ordinary integral over a simultaneity
hyperplane and are preserved under the coordinate transformations. (iii) The
wave equation is linear, of second order at most, generally covariant, and
can be built in a local way with the help of the geometrical objects 
describing the space-time structure. (iv) The probability density 
$\rho = \psi^{*}\psi$ is a scalar field (with the scalar transformation 
rule). In fact the conditions
(i), (ii), (iii) and (iv) are somewhat interrelated.
The coefficients $a, b^{i}, \ldots$
in the wave equation
\begin{displaymath}
\Big[ a\partial_{t}^{2} + b^{i}\partial_{i}\partial_{t} +
c^{ij}\partial_{i}\partial_{j} + f^{i}\partial_{i} + 
d\partial_{t} + g \Big]\psi  = 0,
\end{displaymath}
are local functions of the potential and therefore cannot depend on  
arbitrary high order derivatives of the potential.
Within a rather standard analysis one gets the Schr\"odinger equation 
from (i), (ii), (iii), and (iv), which after the ordinary 
notation of constants has the form
\begin{displaymath} 
\Big[ \frac{\hslash^{2}}{2m}\delta^{ij}\partial_{i}\partial_{j} 
+i\hslash\partial_{t} -m\phi+\Lambda\Big]\psi = 0,
\end{displaymath}
with the $\theta$ in $T_{r}$ given by
\begin{displaymath}
\theta=\frac{m}{2\hslash} \int_{0}^{t} \dot{\vec{A}}^{2}(\tau) \, {\ud{\tau}} 
+\frac{m}{\hslash}\dot{A}_{i}x^{i}.
\end{displaymath} 
$\phi$ is the gravitational potential and $\Lambda$ is one of 
the Kronecker's invariants of the matrix $(\partial_{a}\partial_{b}\phi)$ 
in the above equation.
  
Note that the the inertial mass $m$ in the equation is equal to the parameter 
at the gravitational potential. That is, the gravitational mass must be equal 
to the inertial mass.

\end{document}